\documentclass[11pt,a4paper]{article}
\usepackage{jheppub}
\usepackage{amsmath,amsthm,amssymb}
\usepackage{comment}
\usepackage[dvipsnames,svgnames,x11names]{xcolor}
\usepackage[mathscr]{eucal}
\usepackage{tikz}
\usetikzlibrary{positioning,arrows.meta,calc}
\usepackage{braket}

\usepackage{duckuments}

\usepackage{romannum}
\usepackage{bm}
\usepackage{subfigure}
\usepackage[normalem]{ulem}
\usepackage{mathtools}
\usepackage{stmaryrd}
\makeatletter
\newcommand{\fixed@sra}{$\vrule height 2\fontdimen22\textfont2 width 0pt\shortrightarrow$}
\newcommand{\shortarrow}[1]{%
  \mathrel{\text{\rotatebox[origin=c]{\numexpr#1*45}{\fixed@sra}}}
}
\makeatother
\usepackage[utf8]{inputenc}
\usepackage[T1]{fontenc}
\DeclareMathAlphabet{\mathbbold}{U}{bbold}{m}{n}
\usepackage{hyperref}
\hypersetup{
    colorlinks=true,                 % false: boxed links; true: colored links
    linkcolor=blue,%SeaGreen,                 % color of internal links (change box color with linkbordercolor)
    linktoc=all,                     % link over complete toc entries
    citecolor=SeaGreen,                 % color of links to bibliography
    filecolor=blue,                  % color of file links
    urlcolor=blue                 % color of external links
}
\usepackage{url}
\usepackage{amsthm}
\theoremstyle{plain}
\newtheorem{theorem}{Theorem}
\numberwithin{theorem}{section}
\newtheorem{proposition}[theorem]{Proposition}
\theoremstyle{definition}

\newtheorem*{theorem*}{Theorem}

\newcommand{\defeq}{\mathrel{\mathop:}=}

\DeclareMathOperator{\tr}{tr}
\DeclareMathOperator{\Tr}{Tr}

\newcommand{\Rmnum}[1]{\expandafter\@slowromancap\romannumeral #1@}

\newcommand{\be}{\begin{equation}}
\newcommand{\ee}{\end{equation}}

\newcommand{\C}{\mathbb{C}}

\newcommand{\R}{\mathbb{R}}

\newcommand{\e}{\epsilon}

\renewcommand{\r}{\rho}		%%% Redefinition
\newcommand{\id}{\mathbf 1}
\newcommand{\dd}{\,\mathrm d}
\renewcommand{\e}{\mathrm e}

\title{A density calculus for quantum field theory}

\author[a]{Tanay Kibe}
\emailAdd{tanay.kibe@ib.edu.ar}
\affiliation[a]{Instituto Balseiro, Centro 
At{\'o}mico Bariloche, S.C. de Bariloche, 8400, 
R{\'i}o Negro, Argentina}
\author[b]{and Pratik Roy}
\emailAdd{roy.pratik92@gmail.com}
\affiliation[b]{Institute of Mathematics, University of Warsaw, ul. Banacha 2, 02-097 Warsaw, Poland}

\begin{document}

\abstract{
These notes provide an introduction to Haagerup noncommutative $L^p$ spaces for
quantum field theorists. The aim is to explain how the familiar
calculus of density matrices, fractional powers, and Schatten
norms extends to local observable algebras of type~III, where
an intrinsic trace is unavailable. We develop the construction
first for a matrix algebra, making the crossed product, its
trace, and the dual action explicit, before passing to general
von Neumann algebras. Particular attention is paid to the
representation of normal states by Haagerup densities and to
the distinction between the crossed-product trace and the
canonical Haagerup functional used to compute expectation values.
As a concrete application, we calculate the sandwiched R\'enyi
divergence for a coherent excitation of a chiral current,
verifying monotonicity under restriction and the R\'enyi 
quantum null energy condition at every real order $\alpha>1$.
The presentation assumes familiarity with quantum mechanics
and elementary quantum field theory, while introducing the
operator-algebraic concepts needed for the construction
and calculation.
}

\maketitle

\newpage

\section{Introduction}
\label{sec:introduction}

Quantum information theory provides a useful organizing framework
for many questions in quantum field theory and quantum gravity
\cite{Faulkner:2022mlp}. The quantum null energy condition (QNEC)
and the quantum focusing conjecture (QFC) exemplify the connection
between information, energy, and geometry. QNEC bounds local null
energy in terms of the response of entanglement entropy to a null
deformation of a region
\cite{Bousso:2015wca,Wall:2017blw,Balakrishnan:2017bjg}. In semiclassical gravity,
the QFC proposes a corresponding constraint on null deformations
of generalized entropy, which combines the geometric area of a
surface with the entanglement entropy of quantum fields
\cite{Bousso:2015mna}.

Relative entropy provides a useful way to formulate this
connection. It quantifies the distinguishability between two
quantum states using a specified set of observables. In finite
dimensions, the relative entropy between density matrices $\rho$
and $\sigma$ is
\begin{equation}
  D(\rho\Vert\sigma)
  \coloneqq
  \Tr\!\left[\rho(\log\rho-\log\sigma)\right],
  \label{eq:intro-relative-entropy}
\end{equation}
where $\rho$ is the state of interest and $\sigma$ is the reference
state. It is defined to be $+\infty$ unless the support of $\rho$
is contained in the support of $\sigma$. If $\sigma$ is faithful,
or equivalently full rank, its modular Hamiltonian is
$K_\sigma=-\log\sigma$, and
\begin{equation}
  D(\rho\Vert\sigma)
  =\Delta\langle K_\sigma\rangle-\Delta S,
  \label{eq:intro-relative-entropy-modular-energy}
\end{equation}
where
$\Delta\langle K_\sigma\rangle
\coloneqq\Tr[(\rho-\sigma)K_\sigma]$ and
$\Delta S\coloneqq S(\rho)-S(\sigma)$, with
$S(\rho)\coloneqq-\Tr(\rho\log\rho)$ the von Neumann entropy.
Positivity and monotonicity under quantum channels, including
restriction to a smaller set of observables, turn this identity
into a source of physical constraints. Relative entropy has
accordingly become central to entropy bounds, energy conditions,
modular flow, and bulk--boundary relations in holography
\cite{Casini:2008cr,Wall:2010cj,Blanco:2013joa,Witten:2018zxz,Kibe:2021gtw,Casini:2023Lectures,Iizuka:2025xnd}.
For regions bounded by cuts of a null plane in Minkowski
spacetime, the local QNEC can be expressed as nonnegativity of
the diagonal second null shape variation of relative entropy
with respect to the vacuum \cite{Casini:2017roe,Ceyhan:2018zfg}.

R\'enyi divergences provide a family of further probes of
distinguishability. Their necessity is already apparent for
classical probability distributions. If $p_i$ and $q_i$ are the
probabilities associated with two commuting states, relative
entropy is the mean of the log-likelihood ratio $\log(p_i/q_i)$
in the state $p$. A single mean does not determine the
fluctuations or tails of the random variable. R\'enyi divergences
introduce a parameter $\alpha$ that probes these features with
different weights. For noncommuting density matrices there are
several inequivalent extensions of the classical R\'enyi
divergence. In these notes we focus on the sandwiched R\'enyi
divergence (SRD) \cite{Muller-Lennert:2013liu,Wilde:2013bdg} 
due to its operational significance, its appearance in the 
R\'enyi QNEC discussed presently, as well as its relative simplicity.

In finite dimensions, the SRD
$\widetilde D_\alpha(\rho\Vert\sigma)$ is defined for
$\alpha>0$, $\alpha\ne1$, and recovers relative entropy by
continuous extension at $\alpha=1$. It obeys data processing for
$\alpha\geq\tfrac12$, so restricting access to observables cannot
increase the distinguishability it assigns to two states
\cite{Frank:2013mre,Beigi:2013teh}. We work in this
data-processing range. For faithful finite-dimensional states,
the first correction around $\alpha=1$ is governed by the
relative-entropy variance \cite{Lin:2014azd}. Its distinguished
limits also have direct interpretations: $\alpha=\tfrac12$ is
determined by the fidelity, while $\alpha\to\infty$ is determined
by the smallest constant by which $\sigma$ must be multiplied
to dominate $\rho$ \cite{Tomamichel:2016resources}.

The R\'enyi QNEC asserts that the SRD shares the convexity
of relative entropy under null deformations, requiring its
diagonal second null shape variations to be nonnegative
when the reference state is the vacuum. It was proposed 
in \cite{Lashkari:2018nsl} and has
been established, for $\alpha>1$, in classes of free and
superrenormalizable theories \cite{Moosa:2020jwt,Roy:2022yzm}.
For integer $\alpha\geq2$, an operator-algebraic proof is
available when the deformation is generated by the canonical
translations of a half-sided modular inclusion and the SRD is
finite on the initial, undeformed algebra \cite{Kibe:2026wsg}.
Counterexamples below one occur already in free-field settings
\cite{Moosa:2020jwt}, and extensions involving deformations on
distinct null generators face further obstructions
\cite{Kibe:2026bcn}. These results make precise definitions and
finiteness conditions essential to the study of R\'enyi
divergences in continuum QFT.

The finite-dimensional formulas for relative entropy and the SRD
use traces, density matrices, and fractional powers of density
matrices. Their meaning in continuum QFT requires care.
In algebraic quantum field theory (AQFT), a spacetime region is
associated with a von Neumann algebra of bounded observables
\cite{Fewster:2020Algebraic}. A state assigns expectation values
to these observables, and restriction to a smaller algebra
describes the information accessible in a smaller region.
In broad classes of relativistic QFTs, the relevant local
observable algebras are Type~III factors \cite{Sorce:2023fdx}.
Such algebras admit no faithful normal semifinite trace.
Consequently, there is no intrinsic reduced density matrix in
the local algebra, and restriction cannot be described by the
usual partial trace over a complementary tensor factor.
Relative entropy nevertheless has an intrinsic algebraic
definition, due to Araki \cite{Araki:1976zv}. The challenge is to
recover a similarly useful calculus for the products, fractional
powers, and norms entering more general information measures.

Haagerup noncommutative $L^p$ spaces provide such a calculus
\cite{Haagerup:1979spaces,Terp:1981lp}. They supply canonical
replacements for trace-class density matrices and Schatten
spaces. The construction proceeds through an enlarged algebra,
the crossed product of the original algebra by the modular flow
of a faithful normal state. We work with $\sigma$-finite von
Neumann algebras, meaning that such a state can always be chosen.
The crossed product admits a faithful normal semifinite trace,
denoted by $\tau$, and provides a setting in which the unbounded
operators representing state densities and their powers can be
treated systematically.

The Haagerup $L^p$ spaces consist of suitable operators associated
with the crossed product that obey a prescribed scaling law
under an additional symmetry, the dual action. Normal states
are represented by positive densities $h_\psi\in L^1(M)$,
where $M$ is the original observable algebra. Normalization and
expectation values take the familiar forms
\begin{equation}
  \operatorname{tr}(h_\psi)=1,
  \qquad
  \psi(a)=\operatorname{tr}(h_\psi a),
  \qquad a\in M.
  \label{eq:intro-haagerup-pairing}
\end{equation}
Here $\operatorname{tr}$ is the canonical functional on $L^1(M)$,
distinct from the semifinite trace $\tau$ on the crossed product.
Positive powers of densities belong to the corresponding $L^p$
spaces, and products obey analogues of the familiar H\"older
inequalities and trace identities. Although the construction
begins with a faithful normal state and its modular flow, the
resulting spaces are independent of that choice up to canonical
identifications \cite{Terp:1981lp,Hiai:2021qfv}. This auxiliary
state need not coincide with either state entering a divergence.
The SRD can therefore be formulated intrinsically using these
spaces, with the ordinary matrix expression recovered in finite
dimensions
\cite{Berta:2016vnw,Jencova:2016tqz,Jencova:2017txf,Kato:2023aro,Kato:2023hlj,Hiai:2024qve}.

Crossed product algebras also have an independent physical significance.
In certain semiclassical and holographic settings, they arise
as gravitational observable algebras. An emergent
Type~$\mathrm{III}_1$ algebra can become a
Type~$\mathrm{II}_\infty$ crossed product when gravitational
corrections are included, and its trace allows an entropy to be
defined up to a state-independent additive constant
\cite{Witten:2021unn,Chandrasekaran:2022eqq,Jensen:2023yxy}.
Understanding the crossed product thus also provides a point
of contact with current approaches to perturbative quantum
gravity.

These notes develop the Haagerup density calculus through
explicit examples for quantum field theorists. The monograph
of Goldstein and Labuschagne provides a comprehensive treatment
of noncommutative $L^p$ spaces, including their construction,
interpolation theory, and applications to quantum physics
\cite{Goldstein:2025Noncommutative}. Our emphasis is on a worked
continuum-QFT calculation that connects this general theory to
the SRD and its response to a null displacement. To prepare for
that calculation, we first carry out the Haagerup construction
for a matrix algebra. Although this algebra already has an
ordinary trace, the construction makes explicit how the crossed
product, dual action, and canonical $L^1$ functional recover
familiar density-matrix operations. Realizing the crossed product
as an algebra of matrix-valued functions allows us to identify
the homogeneous operators, recover the Schatten norms, and see
why even a normalized matrix state is represented by an
unbounded Haagerup density.

The field-theory example is the chiral abelian current obtained
from the null-plane correlation functions of a free massive
scalar field. Its half-line algebra is Type~$\mathrm{III}_1$,
and the vacuum modular flow acts by dilations. We construct
the crossed product, evaluate its trace, and exhibit explicit
Haagerup $L^p$ elements. We then prepare a coherent excitation
by a Weyl unitary at an initial cut and, keeping the state
fixed, restrict it to the nested algebras obtained by advancing
the cut. For a particular smearing, a positive-power
factorization and the normalization of an auxiliary state
density yield a closed expression for the SRD at every finite
order $\alpha>1$. Its first and second cut derivatives give
monotonicity under restriction and the R\'enyi QNEC for this
excitation, while the limit $\alpha\to1$ recovers relative
entropy and its coherent-state modular-energy expression.
This example is the main pedagogical focus of the notes,
demonstrating how to calculate with Haagerup densities  
in a local QFT algebra.

The intended readers are graduate students and researchers in
quantum field theory, quantum information, and mathematical
physics. We assume familiarity with Hilbert spaces, density
matrices, and the spectral calculus used in quantum mechanics.
The field-theory example also uses free-field correlation
functions and coherent states. Prior knowledge of von Neumann
algebras, modular theory, or noncommutative $L^p$ spaces is not
required. The concepts needed for the calculations are introduced
as they arise, and general structural theorems are stated with
references to their proofs. Our presentation complements
introductions to operator algebras and modular theory aimed at
physicists
\cite{Witten:2018zxz,Sorce:2023fdx,Sorce:2024pte,Verch:2025Operator}.
For broader accounts of AQFT and entanglement in QFT, see
\cite{Fewster:2020Algebraic,Casini:2023Lectures}, and for an
operator-algebraic treatment of entanglement measures, see
\cite{Hollands:2018Entanglement}. Further references for the
general theory of noncommutative $L^p$ spaces include Terp's
notes, the survey of Pisier and Xu, and Hiai's lectures
\cite{Terp:1981lp,Pisier:2003dsk,Hiai:2021cks}.

The rest of these notes are organized as follows. Section~\ref{sec:finite-dimensional-srd} reviews the
finite-dimensional SRD. Sections~\ref{sec:haagerup-matrices}
and~\ref{sec:algebraic-continuum} develop the Haagerup density
calculus for matrix algebras and general von Neumann algebras,
respectively. Section~\ref{sec:chiral-free-boson} contains the
chiral-current example, and Section~\ref{sec:discussion}
discusses wider applications and further directions.
Appendix~\ref{app:matrix-charts} relates our matrix realization
to the regular crossed-product representation and is not needed
for the main calculations. Readers familiar with the
finite-dimensional SRD may begin with
Section~\ref{sec:haagerup-matrices}; those already comfortable
with Haagerup $L^p$ spaces may consult
Section~\ref{sec:algebraic-continuum} for our conventions before
proceeding to the QFT example.

\section{Finite-dimensional sandwiched R\'enyi divergences}
\label{sec:finite-dimensional-srd}

We begin by reviewing the density matrix calculus that enters the
finite-dimensional SRD. Its definition combines fractional powers of
density matrices with operator multiplication and a Schatten norm.
Understanding how these operations fit together will guide the
construction of the corresponding quantities in continuum QFT.
For a textbook treatment of the finite-dimensional information
measures reviewed here, see \cite{Tomamichel:2016resources}.

\subsection{Density matrices and Schatten norms}
\label{subsec:finite-dimensional-matrix-calculus}

Let \(\mathcal H\) be a Hilbert space of dimension \(d\), and let $B(\mathcal H)$ be the algebra of all bounded linear operators on $\mathcal H$. For an arbitrary matrix \(a \in B(\mathcal H) \), let
\(|a|=(a^*a)^{1/2}\geq0\). Its Schatten \(p\)-norm is
\begin{equation}
  \|a\|_p
  \defeq
  \bigl(\Tr|a|^p\bigr)^{1/p},
  \qquad 0<p<\infty.
  \label{eq:finite-dimensional-schatten-norm}
\end{equation}
It is a norm for \(p\geq1\), and a quasi-norm
for \(0<p<1\), when the usual triangle inequality need not hold. 
Fractional powers of a positive matrix are defined by taking the
corresponding powers of its eigenvalues, using the spectral theorem. 
Positive powers therefore make sense even when some eigenvalues vanish, 
while negative powers require care. The cases \(p=1\) and \(p=2\) 
give the trace norm and Hilbert--Schmidt norm, respectively.
We also write \(\|a\|_\infty\) for the operator norm,
which is the largest singular value. 

The Schatten space, \(S^p(\mathcal H)\),
denotes the matrix space equipped with the corresponding $p$-norm (or
quasi-norm). In finite dimensions all of these spaces contain the same
matrices, and the distinction between the spaces is 
only in how the size of a matrix is measured. 
Denote the set of all positive elements of \(S^p(\mathcal H)\) by \(S^p(\mathcal H)_+\).
Density matrices are  elements of \(S^1(\mathcal H)_+\) with unit
norm. Let
\(\rho,\sigma\in S^1(\mathcal H)_+\) be density matrices,
\begin{equation}
  \rho,\sigma\geq0,
  \qquad
  \Tr\rho=\Tr\sigma=1.
\end{equation}
We regard \(\rho\) as the state whose distinguishability from the reference
state \(\sigma\) is being quantified. Unless stated otherwise, we take
\(\sigma\) to be faithful, which in finite dimensions means that it has
full rank. %All logarithms are natural. 
Our assumption that \(\sigma\) is faithful ensures that all
of its real powers are well defined.

More generally,
\begin{equation}
  \bigl\|\rho^{1/p}\bigr\|_p^p=\Tr\rho=1.
  \label{eq:density-matrix-schatten-normalization}
\end{equation}
In particular, \(\rho^{1/2}\) is a unit vector in the Hilbert space
\(S^2(\mathcal H)\), whose inner product is
\begin{equation}
  \langle a,b\rangle_2=\Tr(a^*b).
  \label{eq:hilbert-schmidt-inner-product}
\end{equation}
Multiplication is controlled by the Schatten version of H\"older's
inequality \cite{Hiai:2021qfv}:
\begin{equation}
  \|ab\|_r\leq\|a\|_p\,\|b\|_q,
  \qquad
  \frac1r=\frac1p+\frac1q,
  \qquad
  1\leq p,q,r\leq\infty,
  \label{eq:schatten-holder}
\end{equation}
where \(1/\infty=0\). For example,
\(\|ab\|_1\leq\|a\|_2\|b\|_2\). This relation between the exponents $p,q,r$
will also organize multiplication in Haagerup \(L^p\) spaces.

\subsection{The sandwiched R\'enyi divergence}
\label{subsec:finite-dimensional-srd-definition}

For classical probability distributions \(p=\{p_i\}\) and \(q=\{q_i\}\),
with \(q_i>0\) whenever \(p_i>0\), the R\'enyi divergence is
\begin{equation}
  D_\alpha(p\Vert q)
  =
  \frac{1}{\alpha-1}
  \log\!\left(\sum_i p_i^\alpha q_i^{1-\alpha}\right),
  \qquad \alpha>0,\quad\alpha\neq1.
  \label{eq:classical-renyi-divergence}
\end{equation}
Equivalently,
\begin{equation}
  D_\alpha(p\Vert q)
  =
  \frac{1}{\alpha-1}
  \log\mathbb E_p
  \!\left[
    \exp\!\left((\alpha-1)\log\frac pq\right)
  \right].
  \label{eq:classical-renyi-cgf}
\end{equation}
Here the log-likelihood ratio is needed only on the support of \(p\). The limit \(\alpha\to1\) recovers the Kullback-Leibler divergence.

For commuting density matrices, a common eigenbasis gives the same
formula. However, for non commuting density matrices multiple extensions are possible depending on the ordering of the fractional powers.
We will focus on one of the many possible choices, the sandwiched R\'enyi divergence, 
which weights \(\rho\) symmetrically
from the left and the right before taking the \(\alpha\)-th power
\cite{Muller-Lennert:2013liu,Wilde:2013bdg}. We will consider the range
\begin{equation}
  \alpha\in
  \left[\frac12,1\right)\cup(1,\infty),
  \label{eq:srd-parameter-range}
\end{equation}
in which the SRD obeys data processing.

Define the
\emph{sandwiched operator}
\begin{equation}
  x_\alpha(\rho\Vert\sigma)
  \defeq
  \sigma^{\frac{1-\alpha}{2\alpha}}
  \rho\,
  \sigma^{\frac{1-\alpha}{2\alpha}}.
  \label{eq:x-alpha}
\end{equation}
It is positive, since
\begin{equation}
  x_\alpha=a_\alpha a_\alpha^*,
  \qquad
  a_\alpha
  \defeq
  \sigma^{\frac{1-\alpha}{2\alpha}}\rho^{1/2}.
  \label{eq:x-alpha-positive}
\end{equation}
Its trace is generally different from one. The sandwiched R\'enyi
moment and divergence are
\begin{align}
  \widetilde Q_\alpha(\rho\Vert\sigma)
  &\defeq
  \Tr\!\left[x_\alpha(\rho\Vert\sigma)^\alpha\right],
  \label{eq:finite-dimensional-srd-Q}\\
  \widetilde D_\alpha(\rho\Vert\sigma)
  &\defeq
  \frac{1}{\alpha-1}
  \log\widetilde Q_\alpha(\rho\Vert\sigma).
  \label{eq:finite-dimensional-srd}
\end{align}
It is easy to see that \eqref{eq:finite-dimensional-srd} recovers \eqref{eq:classical-renyi-divergence} when $\rho$ and $\sigma$ commute.

The essential feature of the SRD for our purposes is its expression
in terms of Schatten norms. Since the eigenvalues of
\(a_\alpha a_\alpha^*\) are the squares of the singular values of
\(a_\alpha\),
\begin{align}
  \widetilde Q_\alpha(\rho\Vert\sigma)
  &=
  \bigl\|x_\alpha(\rho\Vert\sigma)\bigr\|_\alpha^\alpha
  %\notag\\
  %&=
  =\left\|
    \sigma^{\frac{1-\alpha}{2\alpha}}\rho^{1/2}
  \right\|_{2\alpha}^{2\alpha}.
  \label{eq:srd-weighted-schatten}
\end{align}
Consequently,
\begin{equation}
  \widetilde D_\alpha(\rho\Vert\sigma)
  =
  \frac{\alpha}{\alpha-1}
  \log\bigl\|x_\alpha(\rho\Vert\sigma)\bigr\|_\alpha.
  \label{eq:srd-weighted-schatten-divergence}
\end{equation}
The $\alpha$-norm expression in \eqref{eq:srd-weighted-schatten} involves a
quasi-norm when \(\alpha<1\), while the $2\alpha$-norm is an ordinary
Schatten norm throughout the range \(2\alpha\geq1\). The Haagerup construction 
will allow us to naturally generalize these finite-dimensional formulas.

The weighting with respect to the reference state also has a useful 
modular interpretation. With
\(K_\sigma=-\log\sigma\),
\begin{equation}
  x_\alpha(\rho\Vert\sigma)
  =
  \e^{\frac{\alpha-1}{2\alpha}K_\sigma}
  \rho\,
  \e^{\frac{\alpha-1}{2\alpha}K_\sigma}.
  \label{eq:srd-modular-hamiltonian-form}
\end{equation}
For \(\alpha>1\), directions with large \(K_\sigma\), equivalently
small weight in the reference state, are enhanced before the norm is
taken. For \(\alpha<1\), they are suppressed.

\subsection{Distinguished orders}
\label{subsec:finite-dimensional-srd-examples}

Several distinguished orders of the SRD have simple expressions.

\paragraph{Fidelity: \texorpdfstring{\(\alpha=\tfrac12\)}{alpha=1/2}.}
At the lower endpoint, \(x_{1/2}=\sigma^{1/2}\rho\sigma^{1/2}\).
Using the squared-fidelity convention
\begin{equation}
  F(\rho,\sigma)
  \defeq
  \left(
    \Tr\sqrt{\sigma^{1/2}\rho\sigma^{1/2}}
  \right)^2,
  \label{eq:fidelity-convention}
\end{equation}
we obtain
\begin{equation}
  \widetilde D_{1/2}(\rho\Vert\sigma)
  =
  -\log F(\rho,\sigma).
  \label{eq:srd-fidelity-limit}
\end{equation}
Thus this order is determined by the overlap of the two states.

\paragraph{Relative entropy: \texorpdfstring{\(\alpha\to1\)}{alpha to 1}.}
The continuous extension of the SRD at \(\alpha=1\) is the relative entropy,
\begin{equation}
  \widetilde D_1(\rho\Vert\sigma)
  \defeq
  \lim_{\alpha\to1}\widetilde D_\alpha(\rho\Vert\sigma)
  =
  D(\rho\Vert\sigma)
  =
  \Tr(\rho\log\rho-\rho\log\sigma).
  \label{eq:srd-relative-entropy-limit}
\end{equation}
As usual, \(\rho\log\rho\) is defined using \(0\log0=0\).

The way in which this limit is approached contains additional
information. If \(\rho\) as well as \(\sigma\) is faithful, then
\cite{Lin:2014azd}
\begin{equation}
  \widetilde D_\alpha(\rho\Vert\sigma)
  =
  D(\rho\Vert\sigma)
  +
  \frac{\alpha-1}{2}\,V(\rho\Vert\sigma)
  +
  O\!\left((\alpha-1)^2\right),
  \label{eq:srd-near-one-expansion}
\end{equation}
where
\begin{equation}
  V(\rho\Vert\sigma)
  \defeq
  \Tr\!\left[
    \rho\left(
      \log\rho-\log\sigma-D(\rho\Vert\sigma)\id
    \right)^2
  \right]
  \label{eq:relative-entropy-variance}
\end{equation}
is the relative-entropy variance. Relative entropy gives the mean
of \(\log\rho-\log\sigma\) in the state \(\rho\), while the first
derivative of the SRD at \(\alpha=1\) gives one half of its variance.
In the commuting case these are precisely the mean and variance of
the classical log-likelihood ratio.

\paragraph{A Hilbert--Schmidt norm:
\texorpdfstring{\(\alpha=2\)}{alpha=2}.}
At order two,
\begin{align}
  \widetilde D_2(\rho\Vert\sigma)
  &=
  2\log\bigl\|\sigma^{-1/4}\rho\sigma^{-1/4}\bigr\|_2
  \notag\\
  &=
  \log\Tr\!\left(\rho\sigma^{-1/2}\rho\sigma^{-1/2}\right).
  \label{eq:srd-order-two}
\end{align}
The second equality follows by multiplying out the square and using
cyclicity of the trace.

\paragraph{Max-relative entropy:
\texorpdfstring{\(\alpha\to\infty\)}{alpha to infinity}.}
In this limit, \(x_\alpha\) approaches
\(\sigma^{-1/2}\rho\sigma^{-1/2}\), while its Schatten norm tends to
the operator norm. Consequently,
\begin{equation}
  \widetilde D_\infty(\rho\Vert\sigma)
  \defeq
  \lim_{\alpha\to\infty}\widetilde D_\alpha(\rho\Vert\sigma)
  =
  D_{\max}(\rho\Vert\sigma),
  \label{eq:srd-max-limit}
\end{equation}
where
\begin{align}
  D_{\max}(\rho\Vert\sigma)
  &\defeq
  \log\inf\{\lambda>0:\rho\leq\lambda\sigma\}
  \notag\\
  &=
  \log\bigl\|\sigma^{-1/2}\rho\sigma^{-1/2}\bigr\|_\infty.
  \label{eq:max-relative-entropy}
\end{align}
The last equality follows by conjugating the operator inequality
with \(\sigma^{-1/2}\). Thus the upper endpoint is controlled by
the direction in which \(\rho\) is largest relative to \(\sigma\).

\subsection{Structural properties}
\label{subsec:finite-dimensional-srd-properties}

The following properties hold throughout
\eqref{eq:srd-parameter-range}, with the continuous extensions at
\(\alpha=1\) and \(\alpha=\infty\) understood.

\paragraph{Unitary invariance.}
For every unitary \(U\),
\begin{equation}
  \widetilde D_\alpha
  (U\rho U^*\Vert U\sigma U^*)
  =
  \widetilde D_\alpha(\rho\Vert\sigma).
  \label{eq:srd-unitary-invariance}
\end{equation}
Indeed, the sandwiched operator transforms as \(x_\alpha\mapsto
Ux_\alpha U^*\), and its Schatten norm is unchanged.

\paragraph{Tensor-product additivity.}
For two pairs of states,
\begin{equation}
  \widetilde D_\alpha
  (\rho_1\otimes\rho_2\Vert\sigma_1\otimes\sigma_2)
  =
  \widetilde D_\alpha(\rho_1\Vert\sigma_1)
  +
  \widetilde D_\alpha(\rho_2\Vert\sigma_2).
  \label{eq:srd-tensor-additivity}
\end{equation}
The sandwiched operator factorizes, so
\(\widetilde Q_\alpha\) is multiplicative. Taking its logarithm gives
additivity for independent systems.

\paragraph{Data processing.}
If \(\Phi\) is a quantum channel, that is, a completely positive
trace-preserving map, then
\begin{equation}
  \widetilde D_\alpha
  \bigl(\Phi(\rho)\Vert\Phi(\sigma)\bigr)
  \leq
  \widetilde D_\alpha(\rho\Vert\sigma).
  \label{eq:srd-data-processing}
\end{equation}
This is the central structural theorem for the SRD
\cite{Frank:2013mre, Beigi:2013teh}. We quote it here without proof. It says that a
quantum channel cannot increase the distinguishability measured by
the SRD. In particular, discarding part of a bipartite system gives
\begin{equation}
  \widetilde D_\alpha(\rho_{AB}\Vert\sigma_{AB})
  \geq
  \widetilde D_\alpha(\rho_A\Vert\sigma_A).
  \label{eq:srd-partial-trace}
\end{equation}
Here \(\rho_A=\Tr_B\rho_{AB}\), and similarly for \(\sigma_A\).
This is the finite-dimensional precursor of monotonicity under
restriction to a smaller algebra of observables in QFT.

\paragraph{Monotonicity in the order.}
For fixed \(\rho\) and \(\sigma\), the function
\(\alpha\mapsto\widetilde D_\alpha(\rho\Vert\sigma)\) is nondecreasing
\cite{Muller-Lennert:2013liu}. In particular,
\begin{equation}
  -\log F(\rho,\sigma)
  =
  \widetilde D_{1/2}(\rho\Vert\sigma)
  \leq
  D(\rho\Vert\sigma)
  \leq
  D_{\max}(\rho\Vert\sigma).
  \label{eq:srd-order-chain}
\end{equation}

\paragraph{Positivity and definiteness.}
For normalized states, \(0\leq F(\rho,\sigma)\leq1\), with equality
\(F=1\) precisely when the states agree. Monotonicity in the order
therefore gives
\begin{equation}
  \widetilde D_\alpha(\rho\Vert\sigma)\geq0,
  \qquad
  \widetilde D_\alpha(\rho\Vert\sigma)=0
  \quad\Longleftrightarrow\quad
  \rho=\sigma.
  \label{eq:srd-positivity}
\end{equation}

There is an important simplification behind all of the preceding matrix
calculations. A faithful density matrix in finite dimensions has a
strictly positive smallest eigenvalue. Every negative power of
\(\sigma\) is therefore bounded, and the SRD is finite for every
\(\rho\) and every order considered here. Faithfulness alone will
not imply this finiteness in the continuum.

To see how the definition can be adapted, consider \(\alpha>1\)
and rewrite the sandwich as a factorization:
\begin{equation}
  \rho
  =
  \sigma^{c_\alpha}x_\alpha\sigma^{c_\alpha},
  \qquad
  c_\alpha=\frac{\alpha-1}{2\alpha}>0, \quad x_\alpha \in S^\alpha(\mathcal H)_+
  \label{eq:srd-positive-power-factorization}
\end{equation}
This equation characterizes \(x_\alpha\) using only positive powers
of the reference density matrix. Moreover,
\begin{equation}
  2c_\alpha+\frac1\alpha=1.
  \label{eq:srd-factorization-exponents}
\end{equation}
Since \(\sigma^{c_\alpha}\in S^{1/c_\alpha}(\mathcal H)_+\) and
\(x_\alpha\in S^\alpha(\mathcal H)_+\), this is precisely the
H\"older relation that places their product $\rho$ in \(S^1(\mathcal H)_+\).
In the Haagerup formulation, the analogous factorization asks for
an element \(x_\alpha\in L^\alpha\) with finite norm. The existence
of such an element becomes a substantive condition on the two
states.

\section{Haagerup \texorpdfstring{$L^p$}{Lp} spaces for matrices}
\label{sec:haagerup-matrices}

In this section we illustrate the Haagerup \(L^p\)-space construction for matrix
algebras \(M_n(\C)\)  of $n\times n$ complex matrices. Since \(M_n(\C)\) already carries the ordinary matrix trace,
the construction is not needed to define its noncommutative \(L^p\)
spaces. This redundancy is useful because it lets us carry out
the Haagerup construction that will later be applied to a Type~III algebra in
QFT, while checking each step against the familiar Schatten classes.  The discussion in this section expands upon notes by Potapov and Sukochev \cite{Potapo-Sukochev}.

For \(0<p<\infty\), write \(S_n^p\) for \(M_n(\C)\) equipped with the
Schatten norm, or quasi-norm when \(p<1\), introduced in
Section~\ref{sec:finite-dimensional-srd}. The Haagerup $L^p$ spaces for the algebra $M_n(\C)$ is denoted by $L^p(M_n(\C))$. Our first goal is to establish
the isometric identification
\begin{equation}
  J_p:S_n^p\longrightarrow L^p\bigl(M_n(\C)\bigr),
  \qquad
  (J_p(a))(t)=\e^{-t/p}a,
  \label{eq:matrix-Jp-preview}
\end{equation}
where \(t\in\R\) is an auxiliary variable whose significance is explained below. 
We will then follow the
density matrix calculus through this identification. A density matrix $\rho$
will be thought of as a positive \(L^1\) element, with its fractional powers 
$\rho^{1/p}$ belonging to the spaces \(L^p(M_n(\C))\). The construction will also
reproduce expectation values and the finite-dimensional SRD.

\subsection{Modular flow and the crossed product}
\label{subsec:matrix-untwisted-core}
Fix an integer $n$ and denote the corresponding matrix algebra by $M=M_n(\C)$.
Choose a matrix $\varphi \in M_n(\C)$ such that
\begin{equation}
  \varphi>0,
  \qquad
  \Tr \varphi=1.
  \label{eq:matrix-reference-density}
\end{equation}
Since $\varphi$ is strictly positive, it is full rank and defines the faithful state\footnote{In the algebraic language, we will refer to normalized positive functionals on the algebra with sufficient continuity properties as states.}
\begin{equation}
  \omega_\varphi(a)=\Tr(\varphi a),
  \qquad a\in M.
  \label{eq:matrix-reference-state}
\end{equation}
The choice of \(\varphi\) specifies a concrete realization of the Haagerup construction.
We will eventually compare arbitrary states \(\rho\) and \(\sigma\)
within this same realization.

The modular flow of \(\omega_\varphi\) is the unitary conjugation
\begin{equation}
  \sigma_r^\varphi(a)=\varphi^{ir}a\varphi^{-ir},
  \qquad r\in\R.
  \label{eq:matrix-modular-flow}
\end{equation}
For a thermal density matrix \(\varphi=Z^{-1}\e^{-\beta H}\), this is
Hamiltonian evolution with time parameter \(-\beta r\), in the
convention \(a(t)=\e^{itH}a\e^{-itH}\).
More generally, write \(\varphi=\e^{-K}\) and choose a basis in which
\(K=\operatorname{diag}(k_1,\ldots,k_n)\). The matrix units
\(e_{ij}=\ket{i}\!\bra{j}\) then obey
\begin{equation}
  \sigma_r^\varphi(e_{ij})
  =
  \e^{ir(k_j-k_i)}e_{ij}.
  \label{eq:matrix-modular-frequency}
\end{equation}
Thus \(k_j-k_i\) is the modular frequency of \(e_{ij}\).
An additive constant in \(K\) drops out of this flow.

The crossed product is an enlarged algebra containing the original
observables together with a distinguished family of unitaries that
implements their modular evolution. 
To realize it explicitly, introduce the enlarged 
Hilbert space of square integrable vector valued functions,
\begin{equation}
  \widehat{\mathcal H}=L^2(\R_t;\C^n).
  \label{eq:matrix-core-Hilbert-space}
\end{equation}
We label the additional real coordinate in the crossed-product construction by $t$.
Define
\begin{align}
  (\pi(a)\xi)(t)
  &=a\xi(t),
  \label{eq:matrix-untwisted-pi}\\
  (\lambda_\varphi(r)\xi)(t)
  &=\e^{-irt}\varphi^{ir}\xi(t), \quad \xi \in L^2(\mathbb R_t;\mathbb C^n).
  \label{eq:matrix-untwisted-lambda}
\end{align}
Here $\pi(M)$ denotes a representation of the algebra $M$ on the crossed 
product, and the first equation above shows that observables in 
$M$ are represented on the crossed product by constant matrices.
The second equation combines the scalar Fourier multiplier \(\e^{-irt}\)
with the modular unitary \(\varphi^{ir}\). It is easy to verify that
\begin{equation}
  \lambda_\varphi(r)\pi(a)\lambda_\varphi(r)^*
  =
  \pi\bigl(\varphi^{ir}a\varphi^{-ir}\bigr)
  =
  \pi\bigl(\sigma_r^\varphi(a)\bigr).
  \label{eq:matrix-covariance}
\end{equation}
This is the covariance relation for the crossed product, showing that the operators $\lambda_\varphi$ implement modular flow.

These operators give a faithful realization of the crossed product,
whose unitary equivalence with the regular representation commonly encountered in the functional analysis literature is established
in Appendix~\ref{app:matrix-charts}.
We identify the crossed product, also called the \emph{continuous core} 
or simply the \emph{core}, with
\begin{equation}
  \mathcal R_\varphi
  :=
  M\rtimes_{\sigma^\varphi}\R
  =
  \{\pi(M),\lambda_\varphi(r):r\in\R\}''.
  \label{eq:matrix-core-definition}
\end{equation}
The double prime denotes the von Neumann algebra generated by the
listed operators. It includes their linear combinations, products,
adjoints, and weak operator limits. Weak operator convergence means
convergence of matrix elements between every pair of Hilbert-space
vectors.

In this realization both sets of generators act pointwise by matrices 
in $M$, and the generated algebra can be identified explicitly. 
Concretely, since \(\varphi^{-ir}\) belongs to \(M\),
\begin{equation}
  \pi(\varphi^{-ir})\lambda_\varphi(r)
  =
  \e^{-irt}\id_n,
  \label{eq:matrix-scalar-characters}
\end{equation}
where $\id_n\in M$ is the $n\times n$ identity matrix.
The core therefore contains all scalar Fourier multipliers. These
generate \(L^\infty(\R_t)\), while the constant matrix fields% 
\footnote{Here a matrix field is simply a one-parameter family of matrices
$t\mapsto A(t)\in M_n(\mathbb C)$, acting on vector-valued
functions by pointwise multiplication:
$(A\xi)(t)=A(t)\xi(t)$. A constant field is independent of $t$.}
generate \(M_n(\C)\). Hence
\begin{equation}
  \mathcal R_\varphi
  =
  L^\infty(\R_t)\,\overline\otimes\,M_n(\C).
  \label{eq:matrix-core-as-fields}
\end{equation}
The cancellation in \eqref{eq:matrix-scalar-characters} uses the fact
that modular flow is implemented by unitaries inside the matrix algebra.
This is the simplification that makes the matrix example particularly
explicit. Such a simplification will not be available in the QFT case, where the operators $\lambda(r)$ are not in the original algebra $M$.

Elements of \(\mathcal R_\varphi\) can thus be regarded as measurable
matrix fields \(t\mapsto A(t)\) satisfying
\begin{equation}
  \|A\|
  =
  \operatorname*{ess\,sup}_{t\in\R}\|A(t)\|
  <\infty.
  \label{eq:matrix-core-field-norm}
\end{equation}
Measurability means that every matrix entry is measurable. Essential
boundedness means that \(\|A(t)\|\leq C\) almost everywhere in $t$ for some
finite \(C\); the essential supremum is the smallest such bound.
Fields that agree almost everywhere represent the same operator.
Products, adjoints, and positivity are evaluated pointwise in $t$.

The algebra in \eqref{eq:matrix-core-as-fields} does not depend on
\(\varphi\) in this realization. The chosen modular flow is still recorded
by the distinguished unitaries \(\lambda_\varphi(r)\), and will reappear
in the Haagerup density of \(\omega_\varphi\).

\subsection{The dual action and the semifinite trace}
\label{subsec:matrix-dual-action-trace}

The core carries a second action, called the \emph{dual action}.
Here it is simply translation of the auxiliary coordinate:
\begin{equation}
  (\theta_s A)(t)=A(t+s),
  \qquad s\in\R.
  \label{eq:matrix-dual-action}
\end{equation}
It fixes the original observables and multiplies the crossed-product
unitaries by a phase:
\begin{equation}
  \theta_s(\pi(a))=\pi(a),
  \qquad
  \theta_s(\lambda_\varphi(r))
  =
  \e^{-isr}\lambda_\varphi(r).
  \label{eq:matrix-dual-action-generators}
\end{equation}
Conversely, a bounded matrix field fixed by every translation is
constant almost everywhere. The original algebra is therefore exactly
the fixed-point algebra of \(\theta\).

The core has an associated canonical semifinite normal faithful trace of the form
\begin{equation}
  \tau(A)
  =
  \int_\R \e^t\Tr(A(t))\dd t,
  \qquad A\in(\mathcal R_\varphi)_+.
  \label{eq:matrix-core-trace}
\end{equation}
The integral is allowed to be infinite. We use this normalization
throughout the notes.
The trace property follows from cyclicity of the matrix trace.
Faithfulness means that \(\tau(A)=0\) for \(A\geq0\) implies \(A=0\),
which follows from the positivity of the integrand and faithfulness of the matrix trace.
Normality means preservation of increasing positive limits; it follows
from normality of the matrix trace and monotone convergence.
Semifiniteness means that positive elements can be approximated from
below by positive elements of finite trace. For example, consider
\begin{equation}
  E_m(t)=\id_{[-m,m]}(t)\id_n,
  \qquad
  \tau(E_m)=n(\e^m-\e^{-m})<\infty,
  \qquad
  E_m\uparrow\id.
  \label{eq:matrix-finite-trace-exhaustion}
\end{equation}
For \(A\in(\mathcal R_\varphi)_+\), the fields \(E_mAE_m\) increase to \(A\)
and satisfy
\(\tau(E_mAE_m)\leq\|A\|\tau(E_m)<\infty\).

Translation by the dual action rescales the trace:
\begin{align}
  \tau(\theta_s(A))
  &=
  \int_\R\e^t\Tr(A(t+s))\dd t
  \notag\\
  &=
  \e^{-s}\tau(A).
  \label{eq:matrix-core-trace-scaling}
\end{align}
The exponential measure is precisely what turns translation of \(t\)
into this scaling law. This relation between the dual action and the
trace is one of the ingredients that survives in the QFT construction.

\subsection{Homogeneous operators}
\label{subsec:matrix-homogeneous-operators}

The Haagerup construction also uses unbounded operators associated
with the core. Such operators are called \emph{affiliated} operators. 
These operators are unbounded but the spectral projections of their positive parts, and any bounded functions of these operators belong to the core.
In the present realization, they are measurable matrix fields \(A(t)\)
that are finite almost everywhere, but need not be essentially bounded.
They act by multiplication on the vectors \(\xi\) for which
\(\int_\R\|A(t)\xi(t)\|^2\dd t<\infty\).

We need a further condition controlling the large spectral values of
these operators. Define
\begin{equation}
  d_A(\lambda)
  :=
  \tau\!\left(\id_{(\lambda,\infty)}(|A|)\right),
  \qquad \lambda>0.
  \label{eq:matrix-distribution-function}
\end{equation}
At each \(t\), the spectral projection
\(\id_{(\lambda,\infty)}(|A(t)|)\) selects the eigenvalues of
\(|A(t)|\) larger than \(\lambda\). Thus \(d_A(\lambda)\) measures
the size of this large-eigenvalue region using the trace \(\tau\).
The operator is called \(\tau\)-measurable when
\(d_A(\lambda)\to0\) as \(\lambda\to\infty\).

For example, \(A(t)=\e^{-t}\id_n\) is unbounded as \(t\to-\infty\),
but
\begin{equation}
  d_A(\lambda)
  =
  n\int_{-\infty}^{-\log\lambda}\e^t\dd t
  =
  \frac{n}{\lambda}
  \longrightarrow0.
  \label{eq:matrix-scalar-measurability}
\end{equation}
The regions where this operator is large become small when measured
by the trace. We write \(L^0(\mathcal R_\varphi,\tau)\) for the set of all
\(\tau\)-measurable affiliated operators. The dual action extends
to these fields by the same translation formula
\eqref{eq:matrix-dual-action}.

For \(0<p<\infty\), the Haagerup $L^p$ space is defined as
\begin{equation}
  L^p(M)
  =
  \left\{
    A\in L^0(\mathcal R_\varphi,\tau):
    \theta_s(A)=\e^{-s/p}A
    \ \text{for every }s\in\R
  \right\}.
  \label{eq:matrix-haagerup-definition}
\end{equation}
We say that \(A\) has degree \(1/p\) under the dual action. This choice
of exponent ensures that \(|A|^p\) has degree one.

In our realization, the homogeneity condition is
\begin{equation}
  A(t+s)=\e^{-s/p}A(t).
  \label{eq:matrix-homogeneity-equation}
\end{equation}
Equivalently, the measurable field \(t\mapsto\e^{t/p}A(t)\) is
translation invariant, and hence constant almost everywhere.
Using the fact that translation-invariant bounded fields are constant
almost everywhere, there exists an $a\in M_n(\C)$ such that
\[
    e^{(t+s)/p}A(t+s) = e^{t/p}A(t) = a \qquad\text{for almost every }t.
\]
Every solution of this condition has the form
\begin{equation}
  (J_p(a))(t)=\e^{-t/p}a,
  \qquad a\in M_n(\C).
  \label{eq:matrix-homogeneous-solution}
\end{equation}
The remaining question is whether these fields are \(\tau\)-measurable.

Let \(\delta_1,\ldots,\delta_n\) be the singular values of \(a\).
Since
\begin{equation}
  |J_p(a)(t)|=\e^{-t/p}|a|,
\end{equation}
each nonzero singular value contributes to the spectral projection
precisely when
\begin{equation}
  \e^{-t/p}\delta_j>\lambda
  \quad\Longleftrightarrow\quad
  t<p\log\frac{\delta_j}{\lambda}.
\end{equation}
Consequently,
\begin{align}
  d_{J_p(a)}(\lambda)
  &=
  \sum_{\delta_j>0}
  \int_{-\infty}^{\,p\log(\delta_j/\lambda)}
  \e^t\dd t
  \notag\\
  &=
  \lambda^{-p}\sum_{j=1}^n\delta_j^p
  =
  \lambda^{-p}\Tr|a|^p.
  \label{eq:matrix-homogeneous-tail}
\end{align}
This tends to zero as \(\lambda\to\infty\).
The exponential growth of \(J_p(a)\) towards \(t=-\infty\) is
compensated by the exponentially small measure in that direction.
Every field in \eqref{eq:matrix-homogeneous-solution} therefore
belongs to \(L^p(M)\).

For \(p=\infty\), homogeneity means translation invariance, so
\(L^\infty(M)=\pi(M)\), the constant matrix fields. We write
\(J_\infty=\pi\).

\subsection{The Haagerup functional and norm}
\label{subsec:matrix-haagerup-trace-norm}

The tail calculation above establishes measurability, but the core trace
does not give the Haagerup norm by direct integration. Indeed,
for \(a\neq0\),
\begin{equation}
  \tau\bigl(|J_p(a)|^p\bigr)
  =
  \int_\R\e^t
  \Tr\bigl(\e^{-t}|a|^p\bigr)\dd t
  =
  +\infty.
  \label{eq:matrix-core-trace-diverges}
\end{equation}
Thus the homogeneous space \(L^p(M)\) is different from the ordinary
semifinite space \(L^p(\mathcal R_\varphi,\tau)\) of elements with finite $\tau(|\cdot|^p)$.

The Haagerup norm instead uses a functional on \(L^1(M)\), whose elements
are exactly
\begin{equation}
  J_1(b)(t)=\e^{-t}b,
  \qquad b\in M_n(\C).
  \label{eq:matrix-L1-general-form}
\end{equation}
For \(b\geq0\), the tail calculation gives
\begin{equation}
  \tau\!\left(\id_{(\lambda,\infty)}(J_1(b))\right)
  =
  \frac{\Tr b}{\lambda},
  \qquad \lambda>0.
  \label{eq:matrix-L1-tail}
\end{equation}
The finite coefficient \(\Tr b\) is additive on positive matrices
and extends linearly to all matrices. It therefore defines a linear
functional on \(L^1(M)\):
\begin{equation}
  \tr(J_1(b))=\Tr b.
  \label{eq:matrix-haagerup-trace}
\end{equation}
This is the matrix realization of the canonical Haagerup functional
\cite{Terp:1981lp,Hiai:2021qfv}. For positive elements it can equivalently be
read from the spectral tail:
\begin{equation}
  \tr(J_1(b))
  =
  \lambda\,
  \tau\!\left(\id_{(\lambda,\infty)}(J_1(b))\right),
  \qquad b\geq0.
  \label{eq:matrix-haagerup-trace-from-tail}
\end{equation}
Homogeneity and the trace-scaling law explain the independence of
\(\lambda\), and the explicit matrix expression establishes additivity.
In particular, a normalized density matrix has spectral-tail
coefficient one. This fixes the normalization relating \(\tau\)
and \(\tr\).

The two functionals $\tau$ and $\tr$ have different roles in this construction. The core trace \(\tau\)
controls spectral tails and measurability. The functional \(\tr\)
evaluates degree-one operators and supplies the normalization used
for states and norms.
For \(0<p<\infty\) and \(A\in L^p(M)\), define
\begin{equation}
    \|A\|_{L^p(M)}^p=\tr(|A|^p).
\end{equation}
Since
\begin{equation}
  |J_p(a)|^p=J_1(|a|^p),
  \label{eq:matrix-pth-power}
\end{equation}
we obtain
\begin{align}
  \|J_p(a)\|_{L^p(M)}^p
  &=
  \tr\bigl(|J_p(a)|^p\bigr)
  \notag\\
  &=
  \Tr|a|^p,
  \label{eq:matrix-haagerup-schatten-norm}
\end{align}
recovering the Schatten (quasi-)norms on matrices.

\begin{proposition}[Haagerup spaces of a matrix algebra]
\label{prop:matrix-haagerup-schatten}
For every \(0<p<\infty\), the map
\begin{equation}
  J_p:S_n^p\longrightarrow L^p\bigl(M_n(\C)\bigr),
  \qquad
  a\longmapsto\bigl[t\mapsto\e^{-t/p}a\bigr],
  \label{eq:matrix-Jp-isometry}
\end{equation}
is a linear isometric bijection. For \(p<1\), both sides carry their
standard quasi-norms. At \(p=\infty\), the corresponding identification
is the constant embedding \(J_\infty=\pi\).
\end{proposition}

Although the Schatten spaces contain the same matrices for every
\(p\), their Haagerup representatives are different operators. For example
\begin{equation}
  \pi(a)(t)=a,
  \qquad
  J_1(a)(t)=\e^{-t}a,
  \qquad
  J_2(a)(t)=\e^{-t/2}a.
  \label{eq:matrix-different-Lp-copies}
\end{equation}
The \(t\)-dependence records which \(L^p\) space the operator belongs
to. Multiplication and fractional powers will change this dependence
in precisely the way required by the \(L^p\) exponents.

\subsection{Density matrix calculus}
\label{subsec:matrix-density-calculus}

A positive matrix \(\rho\) defines the positive functional
\begin{equation}
  \omega_\rho(a)=\Tr(\rho a),
  \qquad a\in M.
  \label{eq:matrix-functional-rho}
\end{equation}
It is a state when \(\Tr\rho=1\), and normality is automatic in finite
dimensions. Its Haagerup $L^1(M)$ density is
\begin{equation}
  h_\rho\defeq h_{\omega_\rho}=J_1(\rho),
  \qquad
  h_\rho(t)=\e^{-t}\rho.
  \label{eq:matrix-haagerup-density-rho}
\end{equation}
Indeed, \(h_\rho\pi(a)=J_1(\rho a)\), and hence
\begin{equation}
  \tr(h_\rho\pi(a))
  =
  \Tr(\rho a)
  =
  \omega_\rho(a).
  \label{eq:matrix-density-pairing}
\end{equation}
Thus \(h_\rho\) reproduces all expectation values of the original
state. Every positive \(L^1\) element has this form, so states
correspond exactly to positive \(L^1\) elements with \(\tr(h_\rho)=1\).
We will usually suppress \(\pi\) in products, identifying an observable
with its constant matrix field.

Positive fractional powers also have the expected form:
\begin{equation}
  h_\rho^{1/p}=J_p(\rho^{1/p}),
  \qquad
  \|h_\rho^{1/p}\|_{L^p(M)}^p=\Tr\rho,
  \qquad 0<p<\infty.
  \label{eq:matrix-density-roots}
\end{equation}
The first equality follows by taking the power pointwise, and the
second follows from the norm identification. In particular,
\(h_\rho^{1/2}\) is a unit vector in \(L^2(M)\) when \(\rho\) is
normalized. This recovers the square-root description of a density
matrix discussed in Section~\ref{sec:finite-dimensional-srd}.

Multiplication adds the homogeneity degrees. For
\(0<p_1,p_2,p_3\leq\infty\) satisfying
\(1/p_3=1/p_1+1/p_2\), with \(1/\infty=0\), one has
\begin{equation}
  J_{p_1}(a)J_{p_2}(b)=J_{p_3}(ab).
  \label{eq:matrix-degree-addition}
\end{equation}
Adjoints are equally simple:
\begin{equation}
  J_p(a)^*=J_p(a^*).
  \label{eq:matrix-adjoint}
\end{equation}
Both identities follow from pointwise matrix operations.\footnote{Products
of unbounded operators are understood as their closures. In this
realization, the closure is the pointwise matrix product acting on its
natural multiplication domain.}
The Schatten H\"older inequality therefore gives
\begin{align}
  \|J_{p_1}(a)J_{p_2}(b)\|_{L^{p_3}(M)}
  &=
  \|ab\|_{S_n^{p_3}}
  \notag\\
  &\leq
  \|a\|_{S_n^{p_1}}\|b\|_{S_n^{p_2}}
  =
  \|J_{p_1}(a)\|_{L^{p_1}(M)}
  \|J_{p_2}(b)\|_{L^{p_2}(M)}.
  \label{eq:matrix-holder-inequality}
\end{align}
This includes the quasi-norm range \(p_i<1\), with the same exponent
relation \cite{Hiai:2021qfv}.
When \(1/p_1+1/p_2=1\), the products belong to \(L^1(M)\), and
cyclicity takes the familiar form
\begin{align}
  \tr\bigl(J_{p_1}(a)J_{p_2}(b)\bigr)
  &=
  \Tr(ab)
  =
  \Tr(ba)
  \notag\\
  &=
  \tr\bigl(J_{p_2}(b)J_{p_1}(a)\bigr).
  \label{eq:matrix-haagerup-cyclicity}
\end{align}
In particular, the \(L^2\) inner product
\(\langle A,B\rangle_2=\tr(A^*B)\) agrees under \(J_2\) with the
Hilbert--Schmidt inner product.

The same calculus describes a unitary preparation of a state. For
normalized \(\rho\) and a unitary \(U\in M\), let
\(\psi(a)=\omega_\rho(U^*aU)\).
Its density matrix is \(U\rho U^*\), so
\begin{equation}
  h_\psi
  =
  J_1(U\rho U^*)
  =
  U h_\rho U^*.
  \label{eq:matrix-unitary-state-density}
\end{equation}
This identity will be useful later for the coherent states in the QFT example.

For the auxiliary state \(\omega_\varphi\), imaginary powers of its density
recover the crossed-product unitaries:
\begin{equation}
  h_\varphi(t)=\e^{-t}\varphi,
  \qquad
  h_\varphi^{\,ir}(t)
  =
  \e^{-irt}\varphi^{ir}
  =
  \lambda_\varphi(r)(t).
  \label{eq:matrix-reference-density-powers}
\end{equation}
Thus the Haagerup density retains the modular dynamics while 
reproducing all expectation values in the corresponding state.

A faithful state also gives a useful way to produce \(L^p\) elements
from bounded observables. For \(0<p<\infty\), define the symmetric
weighted embedding associated with \(\varphi\) by \cite{KOSAKI198429}
\begin{equation}
  j_p^\varphi(a)
  \defeq
  h_\varphi^{\,1/(2p)}a h_\varphi^{\,1/(2p)}
  =
  J_p\!\left(\varphi^{1/(2p)}a\varphi^{1/(2p)}\right) \in L^p(M).
  \label{eq:matrix-symmetric-embedding}
\end{equation}
Its norm is
\begin{equation}
  \|j_p^\varphi(a)\|_{L^p(M)}
  =
  \left\|\varphi^{1/(2p)}a\varphi^{1/(2p)}\right\|_{S_n^p}.
  \label{eq:matrix-symmetric-embedding-norm}
\end{equation}
This is ordinary symmetric weighting of a matrix followed by \(J_p\).

\subsection{Recovery of the sandwiched R\'enyi divergence}
\label{subsec:matrix-densities-srd}

Let \(\rho\) and \(\sigma\) be density matrices, with \(\sigma>0\),
and keep the auxiliary \(\varphi\) fixed. For \(\alpha\) in the range
\eqref{eq:srd-parameter-range}, write
\(\gamma_\alpha=(1-\alpha)/(2\alpha)\).
The sandwiched matrix \(x_\alpha(\rho\Vert\sigma)\) from
Section~\ref{sec:finite-dimensional-srd} has Haagerup representative
\begin{align}
  X_\alpha
  &=
  J_\alpha\bigl(x_\alpha(\rho\Vert\sigma)\bigr),
  \notag\\
  X_\alpha(t)
  &=
  \e^{-t/\alpha}
  \sigma^{\gamma_\alpha}\rho\sigma^{\gamma_\alpha}.
  \label{eq:matrix-haagerup-sandwich}
\end{align}
This is a positive element of \(L^\alpha(M)\). We now express it
directly in terms of the state densities \(h_\rho\) and \(h_\sigma\).

For \(\tfrac12\leq\alpha<1\), the exponent \(\gamma_\alpha\) is
positive. Since \(2\gamma_\alpha+1=1/\alpha\), multiplication gives
\begin{equation}
  X_\alpha
  =
  h_\sigma^{\,\gamma_\alpha}
  h_\rho
  h_\sigma^{\,\gamma_\alpha}.
  \label{eq:matrix-lower-sandwich}
\end{equation}
Each outer factor belongs to \(L^{1/\gamma_\alpha}(M)\), and the
sum of the three degrees is \(1/\alpha\), as required by the H\"older degree counting \eqref{eq:matrix-holder-inequality}.

For \(\alpha>1\), put
\(c_\alpha=(\alpha-1)/(2\alpha)=-\gamma_\alpha>0\).
A direct sandwich would now involve negative powers of \(h_\sigma\).
Although \(\sigma^{-c}\) is a bounded matrix for every \(c>0\),
the inverse power of the Haagerup density satisfies
\begin{align}
  h_\sigma^{-c}(t)
  &=
  \e^{ct}\sigma^{-c},
  \notag\\
  \tau\!\left(
    \id_{(\lambda,\infty)}(h_\sigma^{-c})
  \right)
  &=
  +\infty,
  \qquad \lambda>0.
  \label{eq:matrix-inverse-not-measurable}
\end{align}
Indeed, all of its eigenvalues exceed any fixed \(\lambda\) for
sufficiently large \(t\), and the measure \(\e^t\dd t\) of that
half-line is infinite. The inverse power is affiliated with the core,
but is not \(\tau\)-measurable.

Pointwise multiplication still evaluates the negative-power sandwich
in this matrix example. To formulate it using positive powers in
the Haagerup spaces, we instead use the factorization introduced
in \eqref{eq:srd-positive-power-factorization}:
\begin{equation}
  h_\rho
  =
  h_\sigma^{\,c_\alpha}
  X_\alpha
  h_\sigma^{\,c_\alpha}.
  \label{eq:matrix-upper-factorization}
\end{equation}
The relation \(2c_\alpha+1/\alpha=1\) places the product in \(L^1(M)\).
For the explicit \(X_\alpha\) above,
\begin{equation}
  \bigl(h_\sigma^{\,c_\alpha}X_\alpha
        h_\sigma^{\,c_\alpha}\bigr)(t)
  =
  \e^{-t}
  \sigma^{c_\alpha+\gamma_\alpha}
  \rho
  \sigma^{\gamma_\alpha+c_\alpha}
  =
  \e^{-t}\rho.
  \label{eq:matrix-upper-factorization-check}
\end{equation}
Thus the factorization holds, and its solution is unique because
\(h_\sigma^{c_\alpha}(t)\) is invertible for every \(t\).

Both ranges of \(\alpha\) therefore lead to the same matrix expression.
Applying the Haagerup functional gives
\begin{align}
  \widetilde Q_\alpha(\omega_\rho\Vert\omega_\sigma)
  &\defeq
  \tr(X_\alpha^\alpha)
  \notag\\
  &=
  \Tr\!\left[
    \left(
      \sigma^{\frac{1-\alpha}{2\alpha}}
      \rho
      \sigma^{\frac{1-\alpha}{2\alpha}}
    \right)^\alpha
  \right].
  \label{eq:matrix-SRD-recovered}
\end{align}
Taking the logarithm,
\begin{equation}
  \widetilde D_\alpha(\omega_\rho\Vert\omega_\sigma)
  \defeq
  \frac{1}{\alpha-1}\log\tr(X_\alpha^\alpha)
  =
  \widetilde D_\alpha(\rho\Vert\sigma).
  \label{eq:matrix-SRD-divergence-recovered}
\end{equation}
The auxiliary choice of \(\varphi\) has dropped out. The result depends
on the two states being compared, with \(\sigma\) entering through
the weighting of \(\rho\).

For matrices, faithfulness of \(\sigma\) guarantees that
\(x_\alpha\) has finite Schatten norm, and hence that
\(X_\alpha\in L^\alpha(M)\). In QFT, the analogous factorization
above one need not have a solution in \(L^\alpha\), and the SRD is then
infinite. Its finiteness will become a substantive constraint on the states in
the field-theory example.

\section{Density matrix calculus for general von Neumann algebras}
\label{sec:algebraic-continuum}

The matrix calculation shows how density matrices, their fractional
powers, and their Schatten norms can be represented by homogeneous
operators in a crossed product. We now explain how the same calculus
extends to local algebras in QFT. The definitions below are the general
versions of those used in Section~\ref{sec:haagerup-matrices}. The main
new ingredient is that modular flow need no longer be implemented by
unitaries belonging to the original algebra. Our purpose  in this section is to collect the ingredients needed for calculations with
states and for the definition of the SRD. We will quote the general
structural results, referring to
\cite{Terp:1981lp, Hiai:2021qfv,Hiai:2021cks} for their proofs. 
In Section~\ref{sec:chiral-free-boson}, these ingredients will be evaluated
explicitly for a local algebra obtained from the free scalar field.

\subsection{From matrix algebras to local algebras}
\label{subsec:local-algebras-reminder}

In algebraic QFT, a spacetime region \(O\) is associated with an algebra
\(\mathcal A(O)\) of bounded observables accessible in that region.
For a real free scalar field $\phi$, one may take \cite{Fewster:2020Algebraic}
\begin{equation}
  \mathcal A(O)
  =
  \left\{
    \e^{i\phi(f)}:
    f\in C_c^\infty(O,\R)
  \right\}'',
  \label{eq:local-Weyl-algebra-reminder}
\end{equation}
where $\phi(f)=\int_{\mathbb R^{1,d-1}}f(x)\phi(x)\dd^d x$ is a smeared field,
and $C_c^\infty(O,\R)$ is the set of all smooth real-valued functions with 
compact support on $O$.
Although the smeared field \(\phi(f)\) is unbounded, its exponential is
a bounded unitary. As in the matrix section, the double prime includes
the weak operator limits needed to obtain the generated von Neumann
algebra. In general, a von Neumann algebra \(M\subset B(\mathcal H)\)
is a unital \(*\)-algebra closed under these limits. Equivalently,
\(M=M''\), where the commutant \(M'\) consists of all bounded operators
commuting with every element of \(M\).

A state specifies expectation values through a linear functional
\(\psi:M\to\C\) satisfying \(\psi(a^*a)\geq0\) and
\(\psi(\id)=1\). For example, the vacuum vector 
$\Omega\in \mathcal H$ gives the state $\omega$, defined by
\(\omega(a)=\langle\Omega,a\Omega\rangle\).
We consider \emph{normal} states, which preserve
increasing bounded limits of positive observables,
\begin{equation}
  0\leq a_i\uparrow a
  \quad\Longrightarrow\quad
  \psi(a_i)\uparrow\psi(a).
  \label{eq:normal-state-reminder}
\end{equation}
Vector states and statistical mixtures of vector states in the chosen
Hilbert-space representation are normal. A state is \emph{faithful}
when \(\psi(a^*a)=0\) implies \(a=0\), generalizing the full-rank
condition on a density matrix. Restricting a state from \(M\) to a
subalgebra \(N\subset M\) simply retains its expectation values on
the observables in \(N\). Recall from above that a trace \(\tau\) 
is \emph{semifinite} when 
every positive observable can be approximated from below by positive
observables with finite trace:
\begin{equation}
    0\leq a_i\uparrow a,
    \qquad \tau(a_i)<\infty.
\end{equation}
Thus the trace need not be finite on every observable,
but finite-trace observables suffice for these approximations.

While admitting state functionals, the Type~III 
local algebras encountered in continuum QFT have no
faithful normal semifinite trace. Thus the formula
\(\psi(a)=\Tr(\rho a)\), with an intrinsic density matrix for the
local algebra, is unavailable. 
The density matrix calculus developed below will represent state
functionals by positive \(L^1\) operators.

For the construction, choose a faithful normal state \(\omega_0\) on
\(M\). An algebra admitting such a state is called
\(\sigma\)-finite.
The subscript distinguishes this auxiliary choice from the reference
state that will enter the SRD. To describe modular flow, we may work
in a representation in which
\(\omega_0(a)=\langle\Omega_0,a\Omega_0\rangle\), with
\(\Omega_0\) cyclic and separating:
\begin{equation}
  \overline{M\Omega_0}=\mathcal H,
  \qquad
  a\Omega_0=0\ \Longrightarrow\ a=0.
  \label{eq:general-cyclic-separating}
\end{equation}
The first (cyclic) condition says that acting with the algebra generates a
dense set of vectors, and the second (separating) ensures faithfulness. Such a
representation is supplied by the GNS construction. In the vacuum
representation of QFT, the Reeh--Schlieder theorem and locality give
these properties for the usual von Neumann algebras of local regions with nonempty causal
complement \cite{Witten:2018zxz,Sorce:2023fdx}. Hence, local algebras in QFT are $\sigma$-finite since the state induced by the cyclic separating vacuum is faithful normal.

Tomita--Takesaki theory associates a positive modular operator
\(\Delta_{\omega_0}\) with this state and algebra.\footnote{For a
concrete definition, let \(S_0\) be the antilinear
operator \(S_0(a\Omega_0)=a^*\Omega_0\). \(S_0\) is well 
defined because \(\Omega_0\) is separating; cyclicity of \(\Omega_0\) ensures a dense
domain for \(S_0\). With \(S\) denoting the closure, the modular operator is defined by
\(\Delta_{\omega_0}=S^*S\). We will use the resulting modular flow
without developing the general theory.}
Its imaginary powers implement the modular flow:
\begin{equation}
  \sigma_r^{\omega_0}(a)
  =
  \Delta_{\omega_0}^{ir}a\Delta_{\omega_0}^{-ir},
  \qquad
  \sigma_r^{\omega_0}(M)=M.
  \label{eq:general-modular-flow}
\end{equation}
The second equality above states that modular flow is an automorphism of the algebra.
For the matrix state \(\omega_D\), the corresponding automorphism is
\(a\mapsto \varphi^{ir}a\varphi^{-ir}\). For a Type~III algebra, modular flow
cannot in general be implemented by a unitary group inside \(M\), and is therefore referred to as an outer automorphism.
The modular operator acts on the representation Hilbert space and
is not a reduced density matrix in the algebra. In sufficiently symmetric QFT
settings the flow nevertheless has a simple geometric meaning. For example it is a Lorentz boost for the vacuum on a Rindler wedge \cite{Bisognano:1975ih}.

\subsection{Haagerup \texorpdfstring{$L^p$}{Lp} spaces and the density matrix calculus}
\label{subsec:haagerup-lp-physics}

The Haagerup $L^p$-space construction uses unbounded operators whose 
spectral data belong to the crossed product, or the core,
\begin{equation}
  \mathcal R
  :=M\rtimes_{\sigma^{\omega_0}}\R.
  \label{eq:haagerup-core}
\end{equation}
As in the matrix example, an untwisted realization acts on
\(\widehat{\mathcal H}=L^2(\R_t;\mathcal H)\) by
\begin{equation}
  (\pi(a)\xi)(t)=a\xi(t),
  \qquad
  (\lambda_{\omega_0}(r)\xi)(t)
  =\e^{-irt}\Delta_{\omega_0}^{ir}\xi(t), \quad a\in M,\;r,t\in \mathbb R.
  \label{eq:general-core-generators}
\end{equation}
These operators generate \(\mathcal R\) and satisfy the covariance relation
\begin{equation}
  \lambda_{\omega_0}(r)\pi(a)\lambda_{\omega_0}(r)^*
  =\pi\bigl(\sigma_r^{\omega_0}(a)\bigr).
  \label{eq:general-core-covariance}
\end{equation}
The modular unitaries now generally lie outside \(M\), so the
cancellation used in \eqref{eq:matrix-scalar-characters} is unavailable.
In particular, the core cannot in general be identified with arbitrary
\(M\)-valued functions of \(t\).

The dual action on the core is characterized by
\begin{equation}
  \theta_s(\pi(a))=\pi(a),
  \qquad
  \theta_s(\lambda_{\omega_0}(r))
  =\e^{-isr}\lambda_{\omega_0}(r).
  \label{eq:general-dual-action}
\end{equation}
It is again implemented by translation of \(t\), and its fixed-point
algebra is \(\pi(M)\). The general crossed-product theorem \cite{Takesaki:duality} guarantees
that \(\mathcal R\) carries a canonical faithful normal semifinite
trace \(\tau\) satisfying
\begin{equation}
  \tau\circ\theta_s=\e^{-s}\tau.
  \label{eq:dual-trace-scaling}
\end{equation}
These are precisely the properties established explicitly for matrices.
They allow us to define spectral tails and measurable operators even
when the original algebra has no trace. From now on we identify
\(M\) with \(\pi(M)\).

A closed densely defined operator is
\emph{affiliated} with an algebra when the spectral projections of
its absolute value and the partial isometry in its polar decomposition
belong to the algebra. For a self-adjoint operator, affiliation is
equivalent to all of its spectral projections belonging to the algebra.
As before, an operator \(A\) affiliated with \(\mathcal{R}\) is \(\tau\)-measurable when
\begin{equation}
  \tau\!\left(\id_{(\lambda,\infty)}(|A|)\right)
  \longrightarrow0
  \qquad\text{as } \lambda\to\infty.
  \label{eq:general-measurability}
\end{equation}
This controls the size, measured by \(\tau\), of the spectral region
where \(|A|\) is large. The measurable affiliated operators form an
algebra \(L^0(\mathcal R,\tau)\), with sums and products understood
as their closures. The dual action extends to this algebra.

The Haagerup $L^p$ spaces are defined to be its homogeneous subspaces:
\begin{equation}
  L^p(M)
  :=
  \left\{
    A\in L^0(\mathcal R,\tau):
    \theta_s(A)=\e^{-s/p}A
    \ \text{for every }s\in\R
  \right\},
  \qquad 0<p\leq\infty,
  \label{eq:haagerup-Lp-definition}
\end{equation}
where \(1/\infty=0\). In particular, \(L^\infty(M)=M\).
This is the same definition used for matrices. 

In the general case,
we no longer solve the homogeneity condition by extracting a constant
matrix coefficient. Instead, the identification of $L^1(M)$ with normal functionals supplies the density operators.
Write \(M_*\) for the space of normal linear functionals on \(M\),
called its predual. Haagerup's construction gives a canonical linear
identification
\begin{equation}
  M_*\simeq L^1(M)
  \label{eq:L1-predual-identification}
\end{equation}
that preserves positivity and norm. A normal positive functional
\(\psi\) therefore corresponds to a unique positive element
\(h_\psi\in L^1(M)\), its \emph{Haagerup density}. The canonical
functional \(\tr:L^1(M)\to\C\) satisfies
\begin{equation}
  \psi(a)=\tr(h_\psi a)=\tr(a h_\psi),
  \qquad a\in M,
  \label{eq:haagerup-density-pairing}
\end{equation}
and hence
\begin{equation}
  \tr(h_\psi)=\psi(\id).
  \label{eq:haagerup-density-normalization}
\end{equation}
States correspond exactly to positive \(L^1\) elements with unit
\(\tr\). Thus \(h_\psi\) contains precisely the expectation-value
data of the state. This is an exact correspondence, valid also when
\(M\) is Type~III.

We reiterate that the distinction between \(\tau\) and \(\tr\) remains essential.
The trace $\tau$ on the core controls measurability, while \(\tr\) evaluates
degree-one operators in \(L^1(M)\). Their relation is the general version of the
matrix spectral-tail formula:
\begin{equation}
  \tau\!\left(\id_{(\lambda,\infty)}(h_\psi)\right)
  =\frac{\psi(\id)}{\lambda},
  \qquad \lambda>0.
  \label{eq:general-density-tail}
\end{equation}
For a nonzero positive functional, \(\tau(h_\psi)=+\infty\), even
though \(\tr(h_\psi)\) is finite. In particular, the Haagerup
density is not a normalized density operator for the core trace.

For \(0<p<\infty\), define the Haagerup $L^p$ norms
\begin{equation}
  \|A\|_p:=\bigl(\tr(|A|^p)\bigr)^{1/p},
  \qquad A\in L^p(M).
  \label{eq:general-haagerup-norm}
\end{equation}
The power \(|A|^p\) has degree one, so the right-hand side is well
defined. These spaces are complete normed spaces for \(p\geq1\)
and quasi-normed spaces for \(0<p<1\). Their norms are intrinsic, i.e.,
changing the auxiliary state \(\omega_0\) gives canonically
isometric spaces, with the same state densities and pairings under
that identification. It does not change any of the information
measures defined below.

Positive powers of a state density obey
\begin{equation}
  h_\psi^{1/p}\in L^p(M),
  \qquad
  \bigl\|h_\psi^{1/p}\bigr\|_p^p=\psi(\id).
  \label{eq:density-root-norm}
\end{equation}
In particular, \(h_\psi^{1/2}\) is a unit vector for a normalized
state. The space \(L^2(M)\) is a Hilbert space with inner product
\(\langle A,B\rangle_2=\tr(A^*B)\), and left multiplication by
\(a\in M\) gives
\begin{equation}
  \langle h_\psi^{1/2},a h_\psi^{1/2}\rangle_2=\psi(a).
  \label{eq:general-L2-state-vector}
\end{equation}
This is the continuum version of representing a density matrix by
its square root in Hilbert--Schmidt space.

Products add degrees and obey the generalized H\"older inequality:
\begin{equation}
  A\in L^{p_1}(M),\quad B\in L^{p_2}(M)
  \quad\Longrightarrow\quad
  AB\in L^{p_3}(M),
  \qquad
  \frac1{p_3}=\frac1{p_1}+\frac1{p_2},
  \label{eq:Lp-degree-addition}
\end{equation}
\begin{equation}
  \|AB\|_{p_3}\leq\|A\|_{p_1}\|B\|_{p_2},
  \qquad 0<p_1,p_2,p_3\leq\infty.
  \label{eq:haagerup-holder}
\end{equation}
The exponent relation is exactly the one seen in the matrix example, and
the inequality also holds in the quasi-norm range. When
\(1/p_1+1/p_2=1\), both products belong to \(L^1(M)\), and
\begin{equation}
  \tr(AB)=\tr(BA).
  \label{eq:general-haagerup-cyclicity}
\end{equation}
These statements include bounded observables by setting one exponent
to infinity, with $1/\infty=0$, justifying the products and cyclic rearrangements used
in the density pairing and the \(L^2\) formula above.

For example, suppose a unitary \(U\in M\) prepares the state
\(\psi(a)=\omega(U^*aU)\). Cyclicity gives
\(\omega(U^*aU)=\tr(Uh_\omega U^*a)\), so uniqueness of the density
implies
\begin{equation}
  h_\psi=U h_\omega U^*.
  \label{eq:general-unitary-state-density}
\end{equation}
This is exactly the usual transformation of a density matrix under
a unitary preparation. We will use this fact later for coherent states of the
scalar field.

Imaginary powers of the Haagerup $L^1$ densities encode the modular flow. For a faithful normal
state \(\omega\),
\begin{equation}
  \sigma_r^\omega(a)=h_\omega^{ir}a h_\omega^{-ir},
  \qquad a\in M,
  \label{eq:modular-flow-haagerup-density}
\end{equation}
where the equality is understood in the core. For the auxiliary
state used in the core construction we get
\begin{equation}
  h_{\omega_0}^{ir}=\lambda_{\omega_0}(r).
  \label{eq:general-auxiliary-density-unitaries}
\end{equation}
Note that the Haagerup density acts on the enlarged Hilbert space, whereas the modular operator
\(\Delta_{\omega_0}\) acts on \(\mathcal H\). They implement the
same automorphisms of the original observables in their respective
representations.

Another useful family of \(L^p\) elements is obtained by symmetric
weighting of bounded observables:
\begin{equation}
  j_p^{\omega_0}(a)
  :=h_{\omega_0}^{1/(2p)}a h_{\omega_0}^{1/(2p)},
  \qquad a\in M.
  \label{eq:general-symmetric-embedding}
\end{equation}
For \(1\leq p<\infty\), its range is dense in \(L^p(M)\), and
H\"older's inequality gives
\(\|j_p^{\omega_0}(a)\|_p\leq\|a\|\)
\cite{KOSAKI198429,Hiai:2021cks}. The weighting depends on
\(\omega_0\), while the space it parametrizes densely is intrinsic
to \(M\).

The resulting dictionary is
\begin{equation}
\begin{array}{c|c}
  \text{matrix algebra } & \text{ general von Neumann algebra}\\[1mm]
  \hline
  \rho\geq0,\ \Tr\rho=1~
    & h_\psi\in L^1(M)_+,\ \tr(h_\psi)=1\\[1mm]
  \Tr(\rho a) & \tr(h_\psi a)=\psi(a)\\[1mm]
  \rho^{1/p}\in S_n^p & h_\psi^{1/p}\in L^p(M)\\[1mm]
  \|a\|_{S_n^p} & \|A\|_p=\tr(|A|^p)^{1/p}
\end{array}
\label{eq:haagerup-physics-dictionary}
\end{equation}
The right-hand column provides a density matrix calculus without a
trace on \(M\). Its density operators belong to the homogeneous
spaces of the core, and its expectation values are evaluated by
\(\tr\).

\subsection{The algebraic sandwiched R\'enyi divergence}
\label{subsec:algebraic-srd}

Let \(\psi\) and \(\omega\) be normal states on \(M\), with
\(\omega\) faithful. We regard \(\psi\) as the state of interest
and \(\omega\) as the reference state. Neither of these states 
needs to coincide with the auxiliary state \(\omega_0\) used to 
construct the core.  We now apply the two forms of the sandwich 
established in the matrix calculation to define the SRD.

For \(\tfrac12\leq\alpha<1\), set
\(\gamma_\alpha=(1-\alpha)/(2\alpha)>0\) and define
\begin{equation}
  X_\alpha(\psi\Vert\omega)
  :=h_\omega^{\gamma_\alpha}h_\psi h_\omega^{\gamma_\alpha}.
  \label{eq:algebraic-lower-sandwich}
\end{equation}
All powers are positive, and H\"older multiplication places this
operator in \(L^\alpha(M)\), since
\begin{equation}
  2\gamma_\alpha+1=\frac1\alpha.
  \label{eq:lower-sandwich-degree}
\end{equation}
It is easy to see that this operator is positive: writing
\(B=h_\psi^{1/2}h_\omega^{\gamma_\alpha}\in L^{2\alpha}(M)\)
gives \(X_\alpha=B^*B\).

For \(\alpha>1\), a direct sandwich would involve negative powers
of \(h_\omega\). Already in the matrix example these inverse powers
were not \(\tau\)-measurable. We therefore use the positive-power
factorization. Put \(c_\alpha=(\alpha-1)/(2\alpha)>0\), and ask
whether there exists \(X_\alpha(\psi\Vert\omega)\in L^\alpha(M)_+\)
such that
\begin{equation}
  h_\psi
  =h_\omega^{c_\alpha}
   X_\alpha(\psi\Vert\omega)
   h_\omega^{c_\alpha}.
  \label{eq:kosaki-factorization-physics}
\end{equation}
The relation
\begin{equation}
  2c_\alpha+\frac1\alpha=1
  \label{eq:upper-factorization-degree}
\end{equation}
makes the right-hand side an \(L^1\) element. Faithfulness of
\(\omega\) ensures uniqueness of the factor \(X_\alpha(\psi\Vert\omega)\) when it exists
\cite{Jencova:2016tqz,Kato:2023aro,Kato:2023hlj}. Its existence is
a further condition on the two states. It requires that the
factor belong to \(L^\alpha(M)\), with finite Haagerup norm.%
\footnote{The positive \(L^1\) elements admitting this factorization
are the positive cone of Kosaki's symmetric space
\(L^\alpha(M,\omega)\). Its norm is the Haagerup \(L^\alpha\)
norm of the factor $X_\alpha$. This connects the definition to the complex interpolation
approach of \cite{KOSAKI198429,Jencova:2016tqz}. Interpolation methods yield useful norm estimates, proofs of
data processing, and monotonicity of the SRD in the R\'enyi
order $\alpha$. They also establish convexity of
$(\alpha-1)\widetilde D_\alpha$ on intervals above one where
it is finite \cite{Berta:2016vnw,Jencova:2016tqz}. We omit these aspects here since the factorization
itself is sufficient for our calculations.}

Whenever \(X_\alpha\) exists, define
\begin{equation}
  \widetilde Q_\alpha(\psi\Vert\omega)
  :=\tr\!\left(X_\alpha(\psi\Vert\omega)^\alpha\right)
  =\bigl\|X_\alpha(\psi\Vert\omega)\bigr\|_\alpha^\alpha.
  \label{eq:QSRD}
\end{equation}
For \(\alpha>1\), if the factorization does not exist in
\(L^\alpha(M)\), set
\begin{equation}
  \widetilde Q_\alpha(\psi\Vert\omega):=+\infty.
  \label{eq:QSRD-infinite}
\end{equation}
The algebraic SRD is then
\begin{equation}
  \widetilde D_\alpha(\psi\Vert\omega)
  :=\frac{1}{\alpha-1}
  \log\widetilde Q_\alpha(\psi\Vert\omega),
  \qquad
  \alpha\in\left[\frac12,1\right)\cup(1,\infty).
  \label{eq:algebraic-SRD}
\end{equation}
It depends only on the algebra and the two states. For a matrix
algebra, the identification \(J_\alpha\) takes the ordinary
sandwiched matrix to \(X_\alpha\), and
\eqref{eq:matrix-SRD-recovered} gives precisely the finite-dimensional
formula. For a Type~III algebra, the same definition uses the
Haagerup densities and their \(L^\alpha\) norms.
As a simple check, when \(\psi=\omega\), both constructions give
\(X_\alpha=h_\omega^{1/\alpha}\). Thus
\(\widetilde Q_\alpha(\omega\Vert\omega)=\tr(h_\omega)=1\)
and \(\widetilde D_\alpha(\omega\Vert\omega)=0\).

This divergence obeys data processing for the displayed range of
\(\alpha\) \cite{Berta:2016vnw,Jencova:2016tqz,Jencova:2017txf}.
In particular, if \(N\subset M\) is a von Neumann subalgebra with
the same identity, then
\begin{equation}
  \widetilde D_\alpha
  \bigl(\psi|_N\Vert\omega|_N\bigr)
  \leq
  \widetilde D_\alpha(\psi\Vert\omega).
  \label{eq:algebraic-SRD-restriction}
\end{equation}
The restricted states are defined by their expectation values on
\(N\), and their densities belong to \(L^1(N)\). For nested
spacetime regions, this inequality says that access to fewer
observables cannot increase distinguishability. It expresses the
information-theoretic content of tracing out a subsystem without
requiring a tensor-product decomposition of the QFT Hilbert space.

In the scalar-field example below, we will take both \(\omega_0\) and
\(\omega\) to be the restricted vacuum state. Its geometric modular
flow will give an explicit core and vacuum density, while a local
Weyl unitary will prepare \(\psi\). The calculation will then
determine the factor \(X_\alpha\), evaluate its norm, and identify
the conditions under which the SRD is finite.

%%%%%%%%%%%%%%%%%%%%%%%%%%%%%%%%%%%%%%%%%%%%%%%%%%%%%%%%%%%%%%
\section{A QFT example: the chiral current}
\label{sec:chiral-free-boson}

We now apply the density matrix calculus of
Section~\ref{sec:algebraic-continuum} to a local QFT algebra.
Our example is a chiral abelian current on a half-line.
Its vacuum is Gaussian and its modular flow is a dilation,
so the ingredients of the Haagerup construction can be
evaluated explicitly. We will construct the algebra and its
core, identify the vacuum density, and then calculate the
SRD of a coherent excitation as the half-line cut is moved.
The calculation will use the current two-point function
together with the pairing, multiplication, and positive-power
rules established in Section~\ref{subsec:haagerup-lp-physics}.

We first explain how this current arises from a free massive
scalar restricted to a null plane, following
\cite{MorinelliTanimotoWegener:2022NullPlane} and
\cite[Section~5]{KibeRoy:2026Quasilocal}. The subsequent
construction concerns one transverse mode of the current
in its vacuum representation.

\subsection{From the massive scalar to null current half-line algebra}
\label{subsec:current-null-plane}

Consider a canonically normalized free real scalar of mass
$m>0$ in $d$-dimensional Minkowski spacetime. Introduce
\begin{equation}
  x^\pm=x^0\pm x^1,\qquad
  \dd s^2=-\dd x^+\dd x^-+\dd y^2,\qquad
  y\in\mathbb R^{d-2}.
  \label{eq:current-null-coordinates}
\end{equation}
On the null plane $x^-=0$, write $u=x^+$. In terms of
$p^\pm=p^0\pm p^1$, the positive mass shell and its measure are
\begin{equation}
  p^+p^-=m^2+|p_\perp|^2,\qquad p^->0,\qquad
  \frac{\dd^{d-1}p}{(2\pi)^{d-1}2p^0}
  =\frac{\dd p^-\dd^{d-2}p_\perp}
         {2p^-(2\pi)^{d-1}}.
  \label{eq:current-null-measure}
\end{equation}
The mass enters the plane wave through $p^+$, conjugate to
$x^-$. On the null plane this part of the phase disappears.
Integrating over transverse momentum therefore gives the
formal vacuum two-point function
\begin{equation}
  \langle\Omega,\phi(u,0,y)\phi(v,0,y')\Omega\rangle
  =\frac{\delta^{d-2}(y-y')}{4\pi}
     \int_0^\infty\frac{\dd k}{k}\,
       \e^{-ik(u-v-i\epsilon)},\qquad
  k=\frac{p^-}{2},\quad \epsilon>0.
  \label{eq:current-formal-scalar-covariance}
\end{equation}
Even though the bulk field is
massive, the null momentum $p^-$ can approach zero, and the above integral diverges at $k=0$.
Consequently, the scalar restricted to the null plane can
only be smeared with functions whose integral along each
null generator vanishes. This removes the zero-momentum
component responsible for the divergence.

Differentiating along the generator supplies the missing
powers of $k$. With the normalization of
\cite{KibeRoy:2026Quasilocal}, define
\begin{equation}
  J(u,y)=\sqrt{2}\,\partial_u\phi(u,0,y).
  \label{eq:current-from-scalar}
\end{equation}
Differentiating \eqref{eq:current-formal-scalar-covariance}
and taking the difference of the two operator orderings yields
\begin{align}
  \langle\Omega,J(u,y)J(v,y')\Omega\rangle
    &=-\frac{\delta^{d-2}(y-y')}{2\pi(u-v-i0)^2},
  \label{eq:current-null-two-point}\\
  [J(u,y),J(v,y')]
    &=i\partial_u\delta(u-v)\,\delta^{d-2}(y-y').
  \label{eq:current-null-commutator}
\end{align}
These distributions are well defined on ordinary test functions,
and the mass has dropped out.

To select a single current, choose a real transverse test
function $e(y)$ with $\int e(y)^2\dd^{d-2}y=1$ and set
$J_e(u)=\int e(y)J(u,y)\dd^{d-2}y$. Then
\begin{equation}
  \langle\Omega,J_e(u)J_e(v)\Omega\rangle
    =-\frac1{2\pi(u-v-i0)^2},\qquad
  [J_e(u),J_e(v)]=i\delta'(u-v).
  \label{eq:current-one-fibre}
\end{equation}
We henceforth suppress $e$ and work in the vacuum Hilbert
space generated by this current. In particular, $\mathcal H$
below contains only the states generated from $\Omega$ by
the current observables. Its higher-point functions are
Gaussian, as inherited from the free scalar.

For a real test function $f\in C_c^\infty(\mathbb R)$, set
\begin{equation}
  J(f)=\int_{\mathbb R}f(u)J(u)\dd u,\qquad
  W(f)=\e^{iJ(f)}.
  \label{eq:current-Weyl}
\end{equation}
There is no zero-average condition on $f$, since integration by
parts gives $J(f)=-\sqrt{2}\phi(f')$, and $f'$ automatically
has zero integral. The smeared current is self-adjoint after
closure, so $W(f)$ is a bounded unitary. As in
\eqref{eq:local-Weyl-algebra-reminder}, these bounded
exponentials will generate the observable algebra.

Smearing the commutator in \eqref{eq:current-one-fibre} gives
\begin{equation}
  [J(f),J(g)]=i\sigma(f,g)\id,\qquad
  \sigma(f,g)=\int_{\mathbb R}f(u)g'(u)\dd u.
  \label{eq:current-symplectic}
\end{equation}
Since the commutator is a scalar, the
Baker--Campbell--Hausdorff (BCH) formula gives
\begin{equation}
  W(f)W(g)=\e^{-i\sigma(f,g)/2}W(f+g),\qquad
  W(f)^*=W(-f).
  \label{eq:current-Weyl-relations}
\end{equation}
Write $\omega(a)=\langle\Omega,a\Omega\rangle$ for vacuum
expectations. Gaussianity means that the variance determines
the Weyl expectation:
\begin{equation}
  \omega(W(f))=\e^{-C_f/2},\qquad
  C_f=\omega(J(f)^2)
     =\frac1{2\pi}\int_0^\infty
        k\left|\int_{\mathbb R}\e^{iku}f(u)\dd u\right|^2\dd k.
  \label{eq:current-Gaussian-vacuum}
\end{equation}
Equations~\eqref{eq:current-Weyl-relations} and
\eqref{eq:current-Gaussian-vacuum} provide the field-theory
input for the calculations below.

Place a cut at $u=V\in\mathbb R$. The algebra of observables
to its right is
\begin{equation}
  M_V=\mathcal A((V,\infty))
   :=\{W(f):f\in C_c^\infty((V,\infty),\mathbb R)\}'',
  \qquad
  \omega_V(a)=\langle\Omega,a\Omega\rangle,\quad a\in M_V.
  \label{eq:current-half-line-algebra}
\end{equation}
The double commutant gives the von Neumann algebra generated
by these unitaries, including their weak operator limits.
$M_V$ is a Type~$\mathrm{III}_1$ factor,
and the vacuum is cyclic and separating for $M_V$ 
\cite{Hollands:2020ChiralModular}. These are the conditions
discussed in \eqref{eq:general-cyclic-separating}:
$M_V\Omega$ is dense in $\mathcal H$, and $a\Omega=0$
implies $a=0$. Thus $\omega_V$ is faithful, since
$\omega_V(a^*a)=\|a\Omega\|^2$, and it is normal because it
is a vector state. It therefore satisfies the hypotheses
of the construction in Section~\ref{sec:algebraic-continuum}.

The general modular flow
\eqref{eq:general-modular-flow} has a geometric interpretation
in this setting. The half-line Bisognano--Wichmann theorem identifies it
with dilations about $V$, obtained by restricting the boosts
illustrated in Figure~\ref{fig:current-moving-cut-setup}
to the null boundary $x^-=0$:
\begin{align}
  \sigma_t^V(J(u))
    &=\e^{-2\pi t}J\bigl(V+\e^{-2\pi t}(u-V)\bigr),
  \notag\\
  \sigma_t^V(W(f))&=W(f_t),\qquad
  f_t(u)=f\bigl(V+\e^{2\pi t}(u-V)\bigr).
  \label{eq:current-cut-modular-flow}
\end{align}
The current formula is understood after smearing. Changing
the integration variable cancels its prefactor and gives
the opposite scaling of the test function
\cite{GarbarzPalau:2023Current}. On the vacuum Hilbert space,
write
\begin{equation}
  \Delta_V^{it}=\e^{-itK_V},\qquad
  K_V=-\log\Delta_V,\qquad K_V\Omega=0.
  \label{eq:current-K}
\end{equation}
$K_V$ is the modular generator for the algebra and vacuum.
Its definition uses the modular operator of
Section~\ref{subsec:local-algebras-reminder} and
does not require a reduced density matrix for the half-line.

\subsection{The core, its trace, and explicit Haagerup elements}
\label{subsec:current-core}
\label{subsec:current-Lp}

We use $\omega_V$ as the auxiliary state in the crossed
product \eqref{eq:haagerup-core}. Substituting its modular
operator into \eqref{eq:general-core-generators} gives
\begin{equation}
  \widehat{\mathcal H}=L^2(\mathbb R_t;\mathcal H),\qquad
  (\pi(a)\xi)(t)=a\xi(t),\qquad
  (\lambda_V(r)\xi)(t)=\e^{-irt}\Delta_V^{ir}\xi(t).
  \label{eq:current-core-generators}
\end{equation}
Here $t$ is the auxiliary coordinate of the core construction.
The operators $\pi(a)$ and $\lambda_V(r)$ generate
$\mathcal R_V=M_V\rtimes_{\sigma^V}\mathbb R$, and their
covariance relation is \eqref{eq:general-core-covariance}.
We identify $M_V$ with its image under $\pi$.

Let $Q$ be multiplication by $t$. Since $Q$ and $K_V$ act
on different tensor factors, they commute. Thus
\begin{equation}
  R_V=K_V+Q,\qquad
  \lambda_V(r)=\e^{-irR_V},\qquad
  h:=h_{\omega_V}=\e^{-R_V},\qquad
  h(t)=\e^{-t}\Delta_V.
  \label{eq:current-vacuum-density}
\end{equation}
The identification of $h$ uses
$h_{\omega_V}^{ir}=\lambda_V(r)$ from
\eqref{eq:general-auxiliary-density-unitaries}.
It is the Haagerup density representing $\omega_V$ through
\eqref{eq:haagerup-density-pairing}. The modular operator
$\Delta_V$ acts on $\mathcal H$, while $h$ acts on the
enlarged space $\widehat{\mathcal H}$.
The spectral projections of $R_V$, and hence of $h$, belong
to the core because its generators include $\e^{-irR_V}$.
This is affiliation in the sense of
Section~\ref{subsec:haagerup-lp-physics}.

For matrices, the modular unitary could be removed from
the crossed-product generator. Here the nontrivial modular
automorphisms are outer, so $\Delta_V^{ir}\notin M_V$ for
$r\ne0$. The cancellation in
\eqref{eq:matrix-scalar-characters} is therefore unavailable.
In particular, $Q$ is not affiliated with the core by itself;
the affiliated combination is $R_V=K_V+Q$.
Although the generators act pointwise in $t$, the core is
not the algebra of arbitrary $M_V$-valued functions of $t$.

The dual action \eqref{eq:general-dual-action} is translation
of $t$. Hence
\begin{equation}
  \theta_s(a)=a,\qquad
  \theta_s(\lambda_V(r))=\e^{-isr}\lambda_V(r),\qquad
  \theta_s(h)=\e^{-s}h.
  \label{eq:current-dual-action}
\end{equation}
The last equality is the degree-one scaling required by
\eqref{eq:haagerup-Lp-definition}.
As shown momentarily, in this representation the canonical core trace is
\begin{equation}
  \tau(X)=\int_{\mathbb R}\e^t
             \langle\Omega,X(t)\Omega\rangle\dd t,\qquad
  X\in(\mathcal R_V)_+.
  \label{eq:current-core-trace}
\end{equation}
Here $X(t)$ denotes the operator acting on $\mathcal H$ at
each value of $t$.
Formula~\eqref{eq:current-core-trace} is the untwisted
representation of the faithful normal semifinite trace
\cite{Witten:2021unn,Hiai:2021qfv}.
The integral is allowed to be infinite.
A change of variable gives
$\tau\circ\theta_s=\e^{-s}\tau$, in agreement with
\eqref{eq:dual-trace-scaling}.

It is instructive to see how a vacuum matrix element
provides a trace. Let $a\in M_V$, $g\in C_c(\mathbb R)$,
and $Y=a g(R_V)$. Since $K_V\Omega=0$,
\begin{equation}
  \tau(Y^*Y)=\omega_V(a^*a)
                \int_{\mathbb R}\e^t|g(t)|^2\dd t.
  \label{eq:current-core-L2-example}
\end{equation}
For the product $YY^*$, use the spectral resolution of
$K_V$ in the vector $a^*\Omega$. If $k$ denotes its spectral
variable, changing variables from $t$ to $r=t+k$ gives
\begin{equation}
  \tau(YY^*)=\|\Delta_V^{1/2}a^*\Omega\|^2
                  \int_{\mathbb R}\e^r|g(r)|^2\dd r
             =\tau(Y^*Y).
  \label{eq:current-trace-check}
\end{equation}
The last step is the Tomita identity
$\|\Delta_V^{1/2}a^*\Omega\|=\|a\Omega\|$,
following from the Tomita operator introduced in
Section~\ref{subsec:local-algebras-reminder}.
Thus the modular factor supplies precisely the correction needed
for cyclicity. 

We can also evaluate the spectral tails explicitly.
Because $h(t)\Omega=\e^{-t}\Omega$, any nonnegative Borel
function $B$ satisfies
\begin{equation}
  \tau(B(h))
   =\int_{\mathbb R}\e^t B(\e^{-t})\dd t
   =\int_0^\infty B(z)\frac{\dd z}{z^2}.
  \label{eq:current-density-spectral-measure}
\end{equation}
In particular,
\begin{equation}
  \tau\!\left(\id_{(a,\infty)}(h)\right)=\frac1a,\qquad
  \tau(h)=+\infty,\qquad \tr(h)=1.
  \label{eq:current-vacuum-tail}
\end{equation}
The first equality verifies measurability by
\eqref{eq:general-measurability}. Together with
\eqref{eq:current-dual-action}, it places $h$ in $L^1(M_V)$.
The normalization $\tr(h)=\omega_V(\id)=1$ is
\eqref{eq:haagerup-density-normalization}, also recovered
from the coefficient of the spectral tail in
\eqref{eq:general-density-tail}.
Thus the infinite core trace $\tau(h)$ is fully consistent
with a normalized Haagerup density.
Likewise, for $0<p<\infty$,
\begin{equation}
  h^{1/p}(t)=\e^{-t/p}\Delta_V^{1/p},\qquad
  \tau\!\left(\id_{(a,\infty)}(h^{1/p})\right)=a^{-p},
  \qquad \|h^{1/p}\|_p=1.
  \label{eq:current-density-roots}
\end{equation}
The degree-$1/p$ scaling and the spectral tail place this
operator in $L^p(M_V)$, realizing the density-root rule
\eqref{eq:density-root-norm} explicitly.

To turn field-theory expectations into Haagerup estimates,
we use the pairing and cyclicity rules of
Section~\ref{subsec:haagerup-lp-physics}.
For $a,b\in M_V$, H\"older multiplication
\eqref{eq:Lp-degree-addition} puts $a h^{1/2}$ and
$b h^{1/2}$ in $L^2(M_V)$, with inner product
\begin{equation}
  \langle a h^{1/2},b h^{1/2}\rangle_2
    =\tr(h^{1/2}a^*b h^{1/2})
    =\tr(h a^*b)=\omega_V(a^*b).
  \label{eq:current-GNS-L2}
\end{equation}
The last two equalities use
\eqref{eq:general-haagerup-cyclicity} and
\eqref{eq:haagerup-density-pairing}, respectively.
Thus $a h^{1/2}\leftrightarrow a\Omega$ preserves inner
products, extending the norm identity
\eqref{eq:general-L2-state-vector}. Completing these dense
sets identifies $L^2(M_V)$ with the vacuum Hilbert space
$\mathcal H$, allowing us to represent physical vectors
as $L^2$ elements when constructing their densities.
The $L^2$ norm uses $\tr$; in particular,
$\|h^{1/2}\|_2=1$ even though $\tau(h)=\infty$.

Vacuum expectations therefore compute Haagerup $L^2$ norms.
For example, taking $a=b=W(f)-\id$ in
\eqref{eq:current-GNS-L2} gives
\begin{equation}
  \|(W(f)-\id)h^{1/2}\|_2^2
     =\omega_V\bigl((W(f)-\id)^*(W(f)-\id)\bigr)
     =2\bigl(1-\e^{-C_f/2}\bigr),
  \label{eq:current-explicit-L2}
\end{equation}
where the last equality follows from
\eqref{eq:current-Gaussian-vacuum}.

To construct the density of the vector functional
defined by $a\Omega$, set $Z=a h^{1/2}$.
H\"older's inequality gives
$ZZ^*=aha^*\in L^1(M_V)_+$, and cyclicity yields
\begin{equation}
  \tr(ZZ^*b)
    =\tr(h a^*ba)
    =\omega_V(a^*ba)
    =\langle a\Omega,b\,a\Omega\rangle,
    \qquad b\in M_V.
  \label{eq:current-vector-density-pairing}
\end{equation}
Hence $ZZ^*$ is its density. Taking $b=\id$ gives
\[
  \tr(ZZ^*)=\|Z\|_2^2
           =\omega_V(a^*a)=\|a\Omega\|^2.
\]
For $a\Omega\neq0$, the normalized vector
$a\Omega/\|a\Omega\|$ therefore defines a state with density
\[
  \frac{ZZ^*}{\|Z\|_2^2}
    =\frac{aha^*}{\omega_V(a^*a)}.
\]
Vacuum expectations thus determine both the normalization
and the observable pairings of the density.

\subsection{Weyl coherent states and their Haagerup densities}
\label{subsec:current-coherent-states}

\begin{figure}[tbp]
\centering
\definecolor{cutblue}{RGB}{67,94,132}
\definecolor{cutteal}{RGB}{32,121,113}

\begin{tikzpicture}[
  x=1cm,y=1cm,font=\small,
  >={Stealth[length=2mm]},
  every node/.style={inner sep=2pt}
]
  \coordinate (A) at (-0.8,-0.8);
  \coordinate (B) at (0.95,0.95);

  % Nested wedges.
  \begin{scope}
    \clip (-1.9,-2.7) rectangle (4.1,3.85);
    \fill[cutblue!10]
      (A) -- (5,5) -- (5,-6.6) -- cycle;
    \fill[cutteal!13]
      (B) -- (5,5) -- (5,-3.1) -- cycle;

    \draw[cutblue!65,dashed,semithick]
      (A) -- (2.2,-3.8);
    \draw[cutteal!65,semithick]
      (B) -- (4.5,-2.6);
  \end{scope}

  % Cartesian axes.
  \draw[->,black!23,thin] (0,-2.6) -- (0,3.65)
    node[above,text=black!50] {$x^0$};
  \draw[->,black!23,thin] (-1.8,0) -- (4.25,0)
    node[right,text=black!50] {$x^1$};

  % Null directions.
  \draw[->,black!55,semithick]
    (2.35,-2.35) -- (-1.6,1.6)
    node[above left,text=black!75] {$x^-$};
  \draw[->,black!45,semithick]
    (-1.85,-1.85) -- (3.95,3.95);

  % Current half-lines on x^-=0.
  \draw[cutblue,line width=1.4pt] (A) -- (B);
  \draw[->,cutteal,line width=1.5pt]
    (B) -- (3.95,3.95);

  \node[anchor=west,align=left] at (4.00,3.95)
    {$u=x^+$\\[-1pt]{\footnotesize $x^-=0$}};

  \filldraw[fill=white,draw=cutblue,line width=1pt]
    (A) circle (2.6pt);
  \filldraw[fill=white,draw=cutteal,line width=1pt]
    (B) circle (2.6pt);

  \node[anchor=south east,text=cutblue]
    at (-0.94,-0.53) {$u=V_0<0$};
  \node[anchor=south east,text=cutteal]
    at (0.82,1.20) {$u=V\geq0$};

  \node[text=cutblue] at (0.85,-1.65) {$u>V_0$};
  \node[text=cutteal] at (3.25,0.25) {$u>V$};

  \path (1.45,1.45) -- (3.35,3.35)
    node[midway,above=5pt,sloped,text=cutteal] {$M_V$};

  \draw[->,cutteal!80!black,line width=0.8pt]
    plot[domain=1.35:-1.1,samples=90,variable=\r]
      ({0.95+1.25*cosh(\r)},{0.95+1.25*sinh(\r)});
  \node[text=cutteal!80!black]
    at (3.10,1.45) {$\sigma_t^V$};
\end{tikzpicture}

\caption{Spacetime section of the null-plane setup, with transverse
directions suppressed. The initial half-line $u>V_0$ (blue and green)
and the later half-line $u>V$ (green) lie on $x^-=0$ and define
$M=M_{V_0}$ and $M_V\subset M$. Shading indicates the corresponding
wedges. The curved arrow shows a vacuum modular orbit in the later
wedge. By the Bisognano--Wichmann theorem, $\sigma_t^V$ acts as a
Lorentz boost about the cut:
$(u-V,x^-)\mapsto(e^{-2\pi t}(u-V),e^{2\pi t}x^-)$.
The arrow points towards increasing $t$. On $x^-=0$, the action
reduces to dilation about $V$.
A Weyl unitary $U=W(f_0)\in M$ prepares the coherent state
$\psi(a)=\omega(U^*aU)$, which is held fixed as the cut advances.
Translation $T_{V-V_0}:M\to M_V$ identifies its restriction with
$\psi_V(a)=\psi(T_{V-V_0}(a))$ on $M$, with the vacuum as reference.}
\label{fig:current-moving-cut-setup}
\end{figure}
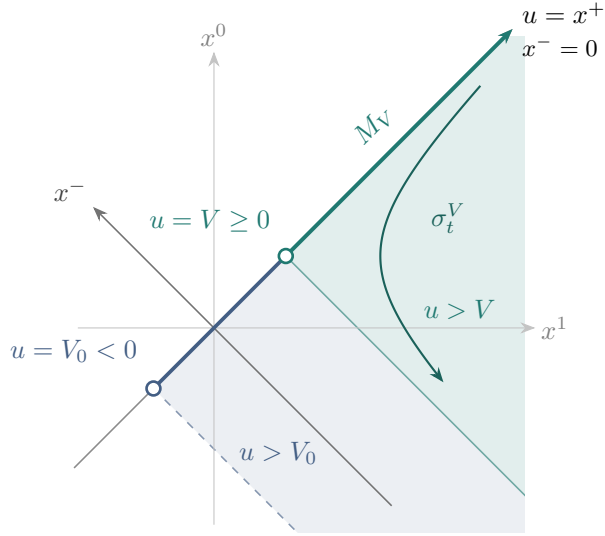

We now construct a coherent excitation and determine the
densities of its restrictions to later half-lines. The 
geometric setup is depicted in Figure~\ref{fig:current-moving-cut-setup}. 
It is convenient to work
throughout in the fixed algebra $M=M_{V_0}$, with $V_0<0$.
In this subsection $\omega=\omega_{V_0}$ and $h=h_\omega$.
We will describe each later restriction as a state on $M$,
so that its density carries the cut dependence.

Choose $\ell>0$ and introduce logarithmic coordinates:
\begin{equation}
  x=u-V_0=\ell\e^r,\qquad
  \mathcal J(r)=xJ(V_0+x),\qquad
  \mathcal J(F)=\int_{\mathbb R}F(r)\mathcal J(r)\dd r,
  \qquad W(F)=\e^{i\mathcal J(F)}.
  \label{eq:current-log-coordinate}
\end{equation}
Thus $\mathcal J(F)=J(f)$ when $F(r)=f(V_0+\ell\e^r)$.
We write $\widehat F(\nu)=\int_{\mathbb R}\e^{-i\nu r}F(r)\dd r$
for the Fourier transform.
% The factor $x$ in $\mathcal J(r)$ absorbs the change of
% integration measure. 
Substituting into
\eqref{eq:current-one-fibre} gives
\begin{equation}
  \omega(\mathcal J(r)\mathcal J(r'))
    =-\frac1{8\pi\sinh^2((r-r'-i0)/2)},
  \label{eq:current-log-covariance}
\end{equation}
making vacuum correlations translation invariant in the logarithmic coordinate $r$.
The Weyl relation, modular flow, and symmetrized covariance become
\begin{align}
  W(F)W(G)&=\e^{-i\Sigma(F,G)/2}W(F+G),
  &\Sigma(F,G)&=\int_{\mathbb R}F(r)G'(r)\dd r,
  \nonumber\\
  h^{it}W(F)h^{-it}&=W(F_t),
  &F_t(r)&=F(r+2\pi t),
  \label{eq:current-log-smearing-norm}\\
  C(F,G)&=\frac1{4\pi}\int_{\mathbb R}
     \nu\coth(\pi\nu)\widehat F(-\nu)\widehat G(\nu)\dd\nu,
  &\omega(W(F))&=\e^{-C_F/2},
  \quad C_F=C(F,F). \nonumber
\end{align}
The modular-flow formula combines the dilation
\eqref{eq:current-cut-modular-flow} with its implementation by
$h^{it}$ in \eqref{eq:modular-flow-haagerup-density}.
Fourier transforming the symmetrized two-point function
\eqref{eq:current-log-covariance} gives the weight
$\nu\coth(\pi\nu)$ in the covariance, with value $1/\pi$ at zero.

The coherent excitation chosen below has a decaying tail rather
than compact support.
To include noncompactly supported smearings, we allow limits of
compactly supported real test functions in the variance norm
$C_F^{1/2}$. Thus we admit a real smearing $F$ whenever there
is a sequence of such test functions $F_n$ with
$C_{F-F_n}\to0$. We must check that this approximation also
defines a Weyl operator belonging to $M$.

The relevant notion of convergence is convergence on each
Hilbert-space vector. We seek an operator $W(F)$ such that
\begin{equation}
  \|W(F_n)\Psi-W(F)\Psi\|\longrightarrow0
  \qquad\text{for every }\Psi.
\end{equation}
This is called \emph{strong operator convergence}. It asks that
the action of the operators converge on each fixed vector,
with no requirement that the convergence be uniform over
all unit vectors. The Gaussian expectation and Weyl relations first show that
$W(F_n)\Psi$ converges when $\Psi$ is a Weyl vector
$W(H)\Omega$, and hence for finite linear combinations of
these vectors. Every Hilbert-space vector can be approximated
by such combinations due to cyclicity of the vacuum. 
If $\Phi$ is a finite linear combination of Weyl vectors
approximating $\Psi$, unitarity gives
\[
  \|(W(F_n)-W(F_m))\Psi\|
  \leq 2\|\Psi-\Phi\|
       +\|(W(F_n)-W(F_m))\Phi\|.
\]
First choose $\Phi$ so that the first term is small.
For this fixed $\Phi$, the convergence already established
makes the second term small when $n,m$ are sufficiently large.
Thus $W(F_n)\Psi$ is a Cauchy sequence and converges for every
vector $\Psi$.
The same argument applies to
$W(F_n)^*=W(-F_n)$, ensuring that the limiting operator $W(F)$
is unitary. Finally, closure under strong operator limits
is a defining property of a von Neumann algebra. Since every
$W(F_n)$ belongs to $M$, so does $W(F)$. We may therefore use
these additional smearings without leaving the observable
algebra.

Choose a real, smooth smearing $f_0$, vanishing near and to the
left of the initial cut $V_0<0$, whose tail is
\begin{equation}
  f_0(u)=\frac{q_0\ell}{u+\ell},\qquad u\geq0,
  \qquad q_0\in\mathbb R.
\end{equation}
The parameter $q_0$ controls the excitation amplitude.
On the interval $(V_0,0)$, we join the zero smearing near $V_0$
smoothly to the prescribed expression near $0$. %, matching all derivatives. 
% The choice $V_0<0$ leaves room for this transition.
Both $f_0$ and its derivative are square integrable, and the
smearing belongs to the finite-variance completion described
above. Thus $W(f_0)$ is a unitary in $M=M_{V_0}$.
It prepares the normal state
\begin{equation}
  \psi(a)=\omega\bigl(W(f_0)^*aW(f_0)\bigr),\qquad
  h_\psi=W(f_0)hW(f_0)^*,\qquad a\in M.
\end{equation}
The second formula is the unitary transformation law
\eqref{eq:general-unitary-state-density}.

To see what this preparation does to the field, conjugate the
current by $W(f_0)$:
\begin{equation}
  W(f_0)^*J(u)W(f_0)=J(u)-f_0'(u)\mathbf1,
  \qquad
  j(u):=\psi(J(u))=-f_0'(u).
  \label{eq:current-coherent-shift}
\end{equation}
As usual, the field identity is understood after smearing.
The preparation therefore adds a classical mean to the vacuum
current. Subtracting that mean leaves all vacuum correlation
functions unchanged. In particular, for $u\geq0$,
\begin{equation}
  j(u)=\frac{q_0\ell}{(u+\ell)^2}.
\end{equation}

We keep this prepared state fixed and restrict it to the later
half-lines with cuts $V\geq0$. 
To compare the restrictions using the same observable algebra,
put $v=V-V_0$ and let $T_v:M\to M_V$ translate observables to
the right by $v$. For example,
\begin{equation}
  T_v\bigl(J(V_0+x)\bigr)=J(V+x).
\end{equation}
We then describe the later restriction as the state on $M$
defined by
\begin{equation}
  \psi_V(a):=\psi(T_v(a)),\qquad a\in M.
  \label{eq:current-fixed-pullback}
\end{equation}
Thus $\psi_V$ assigns to an initial-half-line observable the
expectation of its translated version in the prepared state.
The translation preserves vacuum expectations and is an
isomorphism from $M$ onto $M_V$. Consequently, it also preserves
the SRD when applied to both states:
\begin{equation}
  \widetilde D_\alpha(\psi|_{M_V}\Vert\omega|_{M_V})
  =\widetilde D_\alpha(\psi_V\Vert\omega;M).
\end{equation}
This is the fixed-algebra description of the moving cut used in
\cite[Section~2.3]{Kibe:2026wsg}; see also \cite{Hollands:2025glm}.

Recall that the logarithmic current is
$\mathcal J(r)=xJ(V_0+x)$, with $x=\ell\e^r$.
Denote its expectation in $\psi_V$ by
\begin{equation}
  \chi_V(r):=\psi_V(\mathcal J(r))
  =xj(V+x)
  =\frac{q_0\ell x}{(V+\ell+x)^2}.
\end{equation}
It is useful to express this as a translated profile. Define
\begin{equation}
  A_V=\frac{q_0\ell}{V+\ell},\qquad
  r_V=\log\frac{V+\ell}{\ell},\qquad
  \varphi(r)=\frac{1}{4\cosh^2(r/2)}
            =\frac{\e^r}{(1+\e^r)^2}.
  \label{eq:Av-defn}
\end{equation}
Since $\e^{r-r_V}=x/(V+\ell)$, substitution gives
\begin{equation}
  \chi_V(r)=A_V\varphi(r-r_V),\qquad
  \widehat\chi_V(\nu)
    =\frac{A_V\pi\nu\e^{-i\nu r_V}}{\sinh(\pi\nu)}.
  \label{eq:current-fixed-mean}
\end{equation}

For a real smearing $F$, the expectation of the smeared logarithmic current is
\begin{equation}
  m_V(F):=\psi_V(\mathcal J(F))
  =\int_{\mathbb R}\chi_V(r)F(r)\dd r.
\end{equation}
The centered observable $\mathcal J(F)-m_V(F)\mathbf1$
has the same Gaussian distribution as the vacuum current,
with zero mean and variance $C_F$. Hence, for real $t$,
\[
  \psi_V\!\left(
    \e^{it(\mathcal J(F)-m_V(F)\mathbf1)}
  \right)
  =\e^{-t^2C_F/2}.
\]
Since the subtracted mean is a scalar, restoring it simply
multiplies this expression by $\e^{it m_V(F)}$.
Setting $t=1$ and recalling $W(F)=\e^{i\mathcal J(F)}$, we obtain
\begin{equation}
  \psi_V(W(F))
  =\exp\!\left(
    i\int_{\mathbb R}\chi_V(r)F(r)\dd r-\frac12C_F
  \right).
  \label{eq:current-coherent-characteristic}
\end{equation}
Thus the mean profile and the vacuum covariance
determine all the Weyl expectations of the restricted state.

Note that the original preparation unitary belongs to $M_{V_0}$, so
its existence does not provide a preparation unitary
inside the smaller half-line algebra $M_V$.
We instead identify the density of $\psi_V$ directly by
constructing a positive $L^1(M)$ element with the expectations
\eqref{eq:current-coherent-characteristic}.

For a real smearing $G$ of finite variance, consider the possibly 
unbounded positive vector $\e^{\mathcal J(G)/2}\Omega$.
The action on the vacuum is nevertheless well
defined, since the vacuum distribution of $\mathcal J(G)$ is a
centered Gaussian of variance $C_G$, so
\begin{equation}
  \bigl\|\e^{\mathcal J(G)/2}\Omega\bigr\|^2
  =\omega(\e^{\mathcal J(G)})
  =\e^{C_G/2}<\infty.
\end{equation}
Under the identification of vectors with Haagerup
$L^2$ elements as in \eqref{eq:current-GNS-L2}, this vector
corresponds to
\begin{equation}
  Z_G:=\e^{\mathcal J(G)/2}h^{1/2}\in L^2(M),
  \qquad
  \|Z_G\|_2^2=\e^{C_G/2}.
  \label{eq:current-density-L2-norm}
\end{equation}
To interpret the product when the exponential is unbounded,
first replace it by the bounded spectral truncation
$\e^{\mathcal J(G)/2}\mathbf1_{[-N,N]}(\mathcal J(G))$
and then let $N\to\infty$. The resulting vectors
converge in Hilbert-space norm, and the corresponding products
with $h^{1/2}$ converge in $L^2$.
The product $Z_GZ_G^*$ belongs to $L^1(M)_+$ by H\"older's
inequality. Dividing by its normalization gives
\begin{equation}
  k_G:=\e^{-C_G/2}Z_GZ_G^*
      =\e^{-C_G/2}\e^{\mathcal J(G)/2}h
                            \e^{\mathcal J(G)/2},
  \qquad \tr(k_G)=1,
  \label{eq:current-normalized-Gaussian-density}
\end{equation}
the density of the normalized vector obtained
from $\e^{\mathcal J(G)/2}\Omega$.

We next calculate its Weyl expectations and choose $G$ to
reproduce the desired mean. The quantity $C(G,F)$ introduced
in \eqref{eq:current-log-smearing-norm} is simply the
symmetrized vacuum covariance:
\begin{equation}
  C(G,F)=\frac12\omega\!\left(
     \mathcal J(G)\mathcal J(F)
     +\mathcal J(F)\mathcal J(G)\right).
\end{equation}
In particular, $C(G,G)=C_G$.
Writing $A=\mathcal J(G),\,B=\mathcal J(F)$, and using
\begin{equation}
  \omega\bigl((A+itB)^2\bigr)
  =C_G+2itC(G,F)-t^2C_F.
\end{equation}
one obtains for real $t$,
\begin{equation}
  \bigl\langle\e^{A/2}\Omega,
     \e^{itB}\e^{A/2}\Omega\bigr\rangle
  =\exp\!\left(
     \frac12C_G+itC(G,F)-\frac12t^2C_F\right).
\end{equation}
The vector construction above supplies the meaning of this
matrix element for the unbounded exponentials.
Setting $t=1$ and including the normalization of $k_G$ yields
\begin{equation}
  \tr(k_GW(F))
  =\exp\!\left(-\frac12C_F+iC(G,F)\right).
  \label{eq:current-fixed-Gaussian-source}
\end{equation}
This state has the vacuum variance, and its mean smeared
current is $C(G,F)$.

Comparison with \eqref{eq:current-coherent-characteristic}
shows that we should choose $G$ so that
\begin{equation}
  C(G,F)=\int_{\mathbb R}\chi_V(r)F(r)\dd r
  \qquad\text{for every real test function }F.
\end{equation}
The Fourier formula \eqref{eq:current-log-smearing-norm}
makes this a scalar equation at each frequency. Indeed, matching
\begin{align}
  \int_{\mathbb R}\chi_V(r)F(r)\dd r
    &=\frac1{2\pi}\int_{\mathbb R}
       \widehat\chi_V(-\nu)\widehat F(\nu)\dd\nu.
\end{align}
with \eqref{eq:current-log-smearing-norm} gives
\begin{equation}
  \widehat G_{1,V}(\nu)
  =\frac{2\tanh(\pi\nu)}{\nu}\widehat\chi_V(\nu)
  =\frac{2\pi A_V\e^{-i\nu r_V}}{\cosh(\pi\nu)},
  \qquad
  G_{1,V}(r)=A_V\operatorname{sech}\frac{r-r_V}{2}.
\end{equation}
All quotients at $\nu=0$ are understood by continuity.
This smooth, exponentially decaying profile belongs to the
finite-variance completion. Its variance can be calculated
by setting $F=G_{1,V}$ in the
matching identity to obtain
\begin{align}
  C_{G_{1,V}}
    &=\int_{\mathbb R}\chi_V(r)G_{1,V}(r)\dd r =\frac{A_V^2}{4}\int_{\mathbb R}
       \operatorname{sech}^3\frac{r-r_V}{2}\dd r
      =\frac{\pi A_V^2}{4}.
\end{align}

The normalized density $k_{G_{1,V}}$ now has precisely the
Weyl expectations \eqref{eq:current-coherent-characteristic}.
Agreement on finite linear combinations of Weyl operators now
implies identification of the two normal states on all of $M$.
The uniqueness of the density under
\eqref{eq:L1-predual-identification} gives
$h_{\psi_V}=k_{G_{1,V}}$, or explicitly
\begin{equation}
  h_{\psi_V}=\e^{-\pi A_V^2/8}
      \e^{\mathcal J(G_{1,V})/2}\,h\,
      \e^{\mathcal J(G_{1,V})/2}.
  \label{eq:current-fixed-density}
\end{equation}
The dependence on the later cut is now contained in the density
through $A_V$ and $r_V$, while $M$ and the vacuum density $h$
remain fixed.

\subsection{Sandwiched R\'enyi divergence and null variations}
\label{subsec:current-coherent-SRD}

We now calculate the SRD of the restricted coherent state
relative to the vacuum. We first construct the positive
factor introduced in Section~\ref{subsec:algebraic-srd}
and evaluate its norm, then examine its dependence on the cut.

For $\alpha>1$, put $s=1/\alpha$ and $c=(1-s)/2$.
We seek $X_\alpha(V)\in L^{1/s}(M)_+$ satisfying
\begin{equation}
  h^cX_\alpha(V)h^c=h_{\psi_V}.
  \label{eq:current-SRD-factor-equation}
\end{equation}
The density on the right is the Gaussian sandwich
\eqref{eq:current-fixed-density}.
To see how multiplication by powers of $h$ changes this
expression, differentiate the modular flow
\eqref{eq:current-log-smearing-norm}:
\begin{equation}
  [\log h,\mathcal J(F)]=-2\pi i\mathcal J(F').
  \label{eq:density-commutator}
\end{equation}
Commutators with $\log h$ therefore give linear terms in currents,
while commutators between currents are scalars.
The BCH formula consequently preserves this form, suggesting
that we seek the factor, up to a scalar, among operators
\begin{equation}\label{eq:Ys-defn}
  Y_s(G)=\e^{\mathcal J(G)/2}h^s\e^{\mathcal J(G)/2}.
\end{equation}
Here $G$ is again an unknown real smearing, and $2c+s=1$ fixes
the power of $h$. We will interpret this sandwich as a
fractional power of a normalized density.

We determine $G$ by requiring $h^cY_s(G)h^c$ to reproduce
the density $h_{\psi_V}$ up to an overall scalar.
To compare them, we move all powers of $h$ to the right
and combine the remaining current exponentials.
The preceding commutator formula~\eqref{eq:density-commutator} gives the rule for moving
a power of $h$ past a current. Since
$\widehat{F'}(\nu)=i\nu\widehat F(\nu)$, commutation with
$\log h$ multiplies the smearing coefficient by $2\pi\nu$.
Exponentiating this action gives
\begin{equation}
  h^a\mathcal J(F)h^{-a}=\mathcal J(F^{[a]}),
  \qquad
  \widehat{F^{[a]}}(\nu)
    =\e^{2\pi a\nu}\widehat F(\nu).
\end{equation}
Here $F^{[a]}$ denotes the transformed smearing, which may
be complex. $\mathcal J$ is extended to such smearings by
linearity. These conjugations are first performed with
smooth compact Fourier cutoffs, with the limits taken below.
The corresponding rule for an exponential is
\[
  h^a\e^{\mathcal J(F)/2}
    =\e^{\mathcal J(F^{[a]})/2}h^a.
\]
Applying this rule twice gives
\begin{align}
  h^cY_s(G)h^c
    &=h^c\e^{\mathcal J(G)/2}h^s
        \e^{\mathcal J(G)/2}h^c \notag\\
    &=\e^{\mathcal J(G^{[c]})/2}h^{c+s}
        \e^{\mathcal J(G)/2}h^c \notag\\
    &=\e^{\mathcal J(G^{[c]})/2}
        \e^{\mathcal J(G^{[c+s]})/2}h,
\end{align}
where the last line uses $2c+s=1$.

The commutator of these two currents is a scalar.
The BCH formula therefore combines their exponentials
into a scalar factor times
\[
  \exp\!\left[
    \mathcal J\!\left(\frac{G^{[c]}+G^{[c+s]}}2\right)
  \right].
\]
The smearing in this single current exponential has
Fourier coefficient
\begin{align}
  \frac{\widehat{G^{[c]}}(\nu)
       +\widehat{G^{[c+s]}}(\nu)}2
    &=\frac{\e^{2\pi c\nu}
            +\e^{2\pi(c+s)\nu}}2\,\widehat G(\nu)
      \notag\\
    &=\e^{\pi\nu}\cosh(\pi s\nu)\widehat G(\nu),
\end{align}
again using $2c+s=1$.
This is the coefficient we must match; the scalar BCH
factor will be determined separately.

For the known density $h_{\psi_V}$ in
\eqref{eq:current-fixed-density}, the same rearrangement
has $c=0$, $s=1$, and $G=G_{1,V}$.
Its current coefficient is consequently
\[
  \frac{1+\e^{2\pi\nu}}2\,\widehat G_{1,V}(\nu)
    =\e^{\pi\nu}\cosh(\pi\nu)\widehat G_{1,V}(\nu).
\]
Equating the two coefficients and cancelling
$\e^{\pi\nu}$ determines the required smearing:
\begin{equation}
  \cosh(\pi s\nu)\widehat G_{s,V}(\nu)
    =\cosh(\pi\nu)\widehat G_{1,V}(\nu).
  \label{eq:current-factor-source-matching}
\end{equation}
Substituting the previously obtained expression for $\widehat G_{1,V}(\nu)$ gives
\begin{equation}
  \widehat G_{s,V}(\nu)
    =\frac{2\pi A_V\e^{-i\nu r_V}}{\cosh(\pi s\nu)},
  \qquad
  G_{s,V}(r)=\frac{A_V}{s}
       \operatorname{sech}\frac{r-r_V}{2s}.
  \label{eq:current-factor-source}
\end{equation}
The factorization has therefore fixed both the width and
the amplitude of the smearing.
Only an overall scalar remains to be determined.
Writing $Y_{s,V}=Y_s(G_{s,V})$, the candidate SRD factor is
\begin{equation}
  X_\alpha(V)=\e^{-n_s}Y_{s,V}.
  \label{eq:current-factor-ansatz}
\end{equation}
The scalar $n_s$ accounts for the BCH factors and the
normalization of $h_{\psi_V}$.
We hold $V$ fixed during the calculation.

To compute the norm, we seek a representation
\begin{equation}
  Y_{s,V}=\e^{b_{s,V}}k_{H_{s,V}}^s,
  \qquad
  k_H=\e^{-C_H/2}
       \e^{\mathcal J(H)/2}h\e^{\mathcal J(H)/2}.
  \label{eq:current-factor-density-power}
\end{equation}
The density $k_H$ is normalized by
\eqref{eq:current-normalized-Gaussian-density}, so
\eqref{eq:density-root-norm} gives
$\|k_H^s\|_{1/s}=1$.
Thus $b_{s,V}$ will be the logarithm of the norm.
Writing $G=G_{s,V}$ and $H=H_{s,V}$, we determine both
unknowns from
\begin{equation}
  \log Y_s(G)=b_{s,V}+s\log k_H.
  \label{eq:current-root-log-comparison}
\end{equation}

We first evaluate the logarithm on the left.
Put $A=\mathcal J(G)$ and $L=s\log h$, and write
$DB=[L,B]$ for commutation with $L$.
For example, $D^2A=[L,[L,A]]$.
Equation~\eqref{eq:density-commutator} implies that
$D$ multiplies a current's Fourier smearing coefficient
by $2\pi s\nu$.
Commutators between currents, on the other hand, are
scalars. Consequently, the symmetric BCH expansion
contains only a term linear in the current and a scalar
correction. It can be written as
\begin{equation}
  \log\bigl(\e^{A/2}\e^L\e^{A/2}\bigr)
    =L+f(D)A+[A,g(D)A],
  \label{eq:current-root-symmetric-BCH}
\end{equation}
where the two functions collect the nested commutators,
\begin{equation}
  \begin{aligned}
    f(z)
      &=1+\frac{z^2}{12}-\frac{z^4}{720}+\cdots
        =\frac z2\coth\frac z2,\\
    g(z)
      &=\frac z{24}+\frac{z^3}{360}+\cdots
        =\frac14\coth\frac z2
           \left(\frac z2\coth\frac z2-1\right).
  \end{aligned}
  \label{eq:current-root-BCH-functions}
\end{equation}
Thus the first few terms are
$L+A+D^2A/12+\cdots+[A,DA]/24+\cdots$.
In \eqref{eq:current-root-symmetric-BCH}, functions of $D$
simply act as the corresponding Fourier multipliers.
In particular, the linear current term is
$\mathcal J(P_s)$, with
\begin{equation}
  \widehat P_s(\nu)
    =f(2\pi s\nu)\widehat G(\nu)
    =\pi s\nu\coth(\pi s\nu)\widehat G(\nu).
  \label{eq:current-root-linear-coefficient}
\end{equation}
To evaluate the scalar term, the Weyl relation in
\eqref{eq:current-log-smearing-norm} gives
\[
  [\mathcal J(F),\mathcal J(K)]
    =-\frac1{2\pi}\int_{\mathbb R}
       \nu\,\widehat F(-\nu)\widehat K(\nu)\dd\nu\,\id.
\]
Substituting \eqref{eq:current-root-BCH-functions} into
\eqref{eq:current-root-symmetric-BCH} therefore yields
\begin{equation}
  \log Y_s(G)=s\log h+\mathcal J(P_s)+\frac1{8\pi}\int_{\mathbb R}
       \nu\coth(\pi s\nu)|\widehat G(\nu)|^2\dd\nu
    -\frac{s}{8}\int_{\mathbb R}
       \nu^2\coth^2(\pi s\nu)|\widehat G(\nu)|^2\dd\nu.
  \label{eq:current-root-log-Y}
\end{equation}
The BCH calculations are first performed
with smooth compact Fourier cutoffs.

We now evaluate the right side of
\eqref{eq:current-root-log-comparison}.
By definition,
$\log k_H=\log Y_1(H)-C_H/2$.
Applying \eqref{eq:current-root-log-Y} at $s=1$, its
first scalar integral is exactly $C_H/2$, by
\eqref{eq:current-log-smearing-norm}.
It therefore cancels the normalization term, leaving
\begin{equation}
  b_{s,V}+s\log k_H
    =b_{s,V}+s\log h+s\mathcal J(P_H)
     -\frac{s}{8}\int_{\mathbb R}
       \nu^2\coth^2(\pi\nu)|\widehat H(\nu)|^2\dd\nu,
  \label{eq:current-root-log-k}
\end{equation}
where
$\widehat P_H(\nu)=\pi\nu\coth(\pi\nu)\widehat H(\nu)$.

The terms proportional to $\log h$ already agree.
Matching the current terms in
\eqref{eq:current-root-log-Y} and
\eqref{eq:current-root-log-k} requires $P_s=sP_H$, or
\begin{equation}
  \coth(\pi\nu)\widehat H(\nu)
    =\coth(\pi s\nu)\widehat G(\nu).
  \label{eq:current-root-current-matching}
\end{equation}
Using \eqref{eq:current-factor-source}, we obtain
\begin{equation}
  \widehat H_{s,V}(\nu)
    =\frac{\tanh(\pi\nu)}{\tanh(\pi s\nu)}
       \widehat G_{s,V}(\nu)
    =\frac{2\pi A_V\e^{-i\nu r_V}\tanh(\pi\nu)}
           {\sinh(\pi s\nu)}.
  \label{eq:current-factor-root-source}
\end{equation}
The value at zero is understood by continuity.
This transform is bounded near zero and decays
exponentially at infinity, so $C_{H_{s,V}}<\infty$
and the normalized density $k_{H_{s,V}}$ is well defined.

Equation~\eqref{eq:current-root-current-matching} also
makes the last scalar integrals in
\eqref{eq:current-root-log-Y} and
\eqref{eq:current-root-log-k} equal.
Comparing the remaining scalar terms therefore fixes
\begin{align}
  b_{s,V}
    &=\frac1{8\pi}\int_{\mathbb R}
       \nu\coth(\pi s\nu)
       |\widehat G_{s,V}(\nu)|^2\dd\nu \notag\\
    &=2\pi A_V^2\int_0^\infty
       \frac{\nu\dd\nu}{\sinh(2\pi s\nu)}
      =\frac{\pi A_V^2}{8s^2}.
  \label{eq:current-factor-root-scalar}
\end{align}
The second line follows by substituting
\eqref{eq:current-factor-source} and using
$\int_0^\infty x/\sinh x\,\dd x=\pi^2/4$.
At $s=1$, we recover $H_{1,V}=G_{1,V}$ and
$b_{1,V}=C_{G_{1,V}}/2$.

We have now established
\eqref{eq:current-factor-density-power}, including its
normalization. Since $k_{H_{s,V}}$ is a normalized
$L^1$ density, this gives
\begin{equation}
  Y_{s,V}\in L^{1/s}(M)_+,
  \qquad
  \log\|Y_{s,V}\|_{1/s}=b_{s,V},
  \qquad
  \log\tr(Y_{s,V}^{1/s})
    =\frac{b_{s,V}}s=\frac{\pi A_V^2}{8s^3}.
  \label{eq:current-fixed-root-moment}
\end{equation}
The finite covariance integrals allow the Fourier cutoffs
to be removed in the Gaussian vectors and densities.
Continuity of positive fractional powers then gives the
corresponding $L^{1/s}$ limits, and H\"older multiplication
controls the products with $h^c$.

It remains to determine $n_s$.
The source matching in
\eqref{eq:current-factor-source-matching} ensures that
$h^cY_{s,V}h^c$ is proportional to $h_{\psi_V}$.
Since the latter density has unit trace,
\begin{equation}
  n_s=\log\tr(h^cY_{s,V}h^c),
  \qquad
  h^cY_{s,V}h^c=\e^{n_s}h_{\psi_V}.
  \label{eq:current-fixed-compression}
\end{equation}
To evaluate this trace, continue to write $G=G_{s,V}$.
Cyclicity, $2c=1-s$, and
\eqref{eq:density-commutator} give
\begin{equation}
  \tr(h^cY_s(G)h^c)
    =\omega\!\left(
       \e^{\mathcal J(G^{[-s]})/2}
       \e^{\mathcal J(G)/2}\right),
  \qquad
  \widehat{G^{[-s]}}(\nu)
    =\e^{-2\pi s\nu}\widehat G(\nu).
  \label{eq:current-compression-expectation}
\end{equation}
At the cutoff level, the Gaussian expectation in
\eqref{eq:current-log-smearing-norm} and the scalar
current commutator imply
\[
  \log\omega\!\left(
    \e^{\mathcal J(F)/2}\e^{\mathcal J(G)/2}\right)
    =\frac18\bigl(C_F+C_G+2C(F,G)\bigr)
     +\frac18[\mathcal J(F),\mathcal J(G)].
\]
Applying this identity to
\eqref{eq:current-compression-expectation}, and combining
the scalar terms before removing the cutoffs, gives
\begin{align}
  n_s
    &=\frac1{16\pi}\int_{\mathbb R}
       \nu\,
       \frac{\cosh(\pi\nu)+\cosh((1-2s)\pi\nu)}
            {\sinh(\pi\nu)}
       |\widehat G_{s,V}(\nu)|^2\dd\nu \notag\\
    &=\pi A_V^2\int_0^\infty
       \nu\bigl[\coth(\pi\nu)-\tanh(\pi s\nu)\bigr]\dd\nu
       \notag\\
    &=\frac{\pi A_V^2}{24}(s^{-2}+2).
  \label{eq:current-compression-scalar}
\end{align}

Substituting \eqref{eq:current-compression-scalar} into
\eqref{eq:current-factor-ansatz} gives
\begin{equation}
  X_\alpha(V)=
    \exp\!\left[-\frac{\pi A_V^2}{24}(\alpha^2+2)\right]
    \e^{\mathcal J(G_{1/\alpha,V})/2}
    h^{1/\alpha}
    \e^{\mathcal J(G_{1/\alpha,V})/2}.
  \label{eq:current-fixed-factor}
\end{equation}
Equation~\eqref{eq:current-fixed-root-moment} places this
positive operator in $L^\alpha(M)$, and
\eqref{eq:current-fixed-compression} gives the required
factorization \eqref{eq:current-SRD-factor-equation}.
Faithfulness of $\omega$ and the uniqueness established
in Section~\ref{subsec:algebraic-srd} therefore identify
\eqref{eq:current-fixed-factor} as the SRD factor.

Its moment follows directly from
\eqref{eq:current-factor-density-power},
\eqref{eq:current-factor-root-scalar}, and
\eqref{eq:current-compression-scalar},
\begin{equation}
  \log\tr(X_\alpha^\alpha)
    =\frac{b_{s,V}-n_s}{s}
    =\frac{\pi A_V^2}{12}(\alpha^3-\alpha).
  \label{eq:current-SRD-moment}
\end{equation}
Using \eqref{eq:QSRD}, \eqref{eq:algebraic-SRD}, and
\eqref{eq:Av-defn}, we obtain
\begin{equation}
  \widetilde D_\alpha(V)
    =\frac{\pi\alpha(\alpha+1)q_0^2\ell^2}
           {12(V+\ell)^2},
  \qquad \alpha>1,\quad V\geq0.
  \label{eq:current-explicit-SRD}
\end{equation}
The phase $\e^{-i\nu r_V}$ cancels from the scalar
integrals, so the cut dependence enters only through
$A_V^2$.

Differentiating \eqref{eq:current-explicit-SRD} gives
\begin{align}
  \partial_V\widetilde D_\alpha
    &=-\frac{\pi\alpha(\alpha+1)q_0^2\ell^2}
             {6(V+\ell)^3}\leq0,
      \notag\\
  \partial_V^2\widetilde D_\alpha
    &=\frac{\pi\alpha(\alpha+1)q_0^2\ell^2}
            {2(V+\ell)^4}\geq0.
\end{align}
These derivatives compare restrictions of one fixed
prepared state to nested half-lines.
The nonpositive first derivative agrees with data
processing, \eqref{eq:algebraic-SRD-restriction}.
The nonnegative second derivative verifies the
R\'enyi QNEC for this excitation at every real
$\alpha>1$.

Finally, taking $\alpha\to1$ in
\eqref{eq:current-explicit-SRD} gives the relative entropy
\begin{equation}
  D_1(V)
    =\frac{\pi q_0^2\ell^2}{6(V+\ell)^2}
    =\pi\int_V^\infty(u-V)j(u)^2\dd u,
  \label{eq:current-relative-entropy-limit}
\end{equation}
in agreement with \cite{Longo:2020CurrentEntropy}.
Differentiating the integral twice gives
$D_1''(V)=\pi j(V)^2$.
The coherent state has the same centered two-point
function as the vacuum, so vacuum normal ordering in
$T(u)=\tfrac12:J(u)^2:$ leaves only the squared mean,
\begin{equation}
  \Delta\langle T(u)\rangle=\frac12j(u)^2,
  \qquad
  D_1''(V)=\pi j(V)^2
         =2\pi\Delta\langle T(V)\rangle.
  \label{eq:current-coherent-energy}
\end{equation}
Thus the second cut derivative of relative entropy
recovers the local excess null energy density.

\section{Discussion}
\label{sec:discussion}

We have attempted to provide a practical introduction to Haagerup
$L^p$ spaces through the density-matrix operations familiar from
quantum mechanics. The matrix example makes the crossed product,
dual action, and canonical trace functional explicit while recovering
the Schatten calculus. We then extended this description to
general von Neumann algebras and applied it to the
Type~$\mathrm{III}_1$ algebra of a chiral current on a half-line.
For a Weyl excitation, expressing the SRD factor as a scalar
multiple of a fractional power of a normalized auxiliary
density made its norm explicit.
The resulting SRD gives monotonicity under restriction
and the R\'enyi QNEC for this excitation at every real
$\alpha>1$.

The construction provides a density calculus, 
closely paralleling tracial density matrix calculus, even when
the local observable algebra has no trace. Normal states have
positive $L^1$ densities, their square roots belong to $L^2$,
and H\"older inequalities control their products. The worked
example also distinguishes preparation from restriction.
Although the state is prepared by a Weyl unitary in the initial
algebra, its restrictions need not admit such a preparation
inside the smaller algebras. Restriction preserves normality,
so these states nevertheless continue to admit Haagerup
densities. The calculus therefore remains applicable as the
accessible observables change.

Further explicit calculations for non-coherent states would
clarify how far the factorization used here extends.
Physical criteria for finiteness of higher-order divergences
and applications to regions whose modular flow is not
geometric are also useful directions. In each case, the
intrinsic $L^p$ calculus supplies a concrete setting for the calculation.
The remaining task is to obtain sufficient information about
the states and modular dynamics of the theory.

A powerful tool, not developed in these notes, is
\emph{complex interpolation}. Roughly speaking, an estimate
known at two values of $p$ can yield estimates at the
intermediate values. For example, if a linear map does not
increase norms at either endpoint, interpolation can show
that it does not increase the intermediate norms either.
To make this comparison precise, one uses a faithful reference
state to relate the different Haagerup $L^p$ spaces
\cite{KOSAKI198429,Jencova:2016tqz}.
The word \emph{complex} refers to the use of analytic functions
of a complex parameter to obtain these intermediate bounds
\cite{bergh1976interpolation}.
This provides a systematic way to prove norm inequalities,
contractivity, and convexity properties. Applied to suitable weighted density operators, interpolation
shows that, for fixed states,
\begin{equation}
  \alpha\longmapsto
  \log\widetilde Q_\alpha(\psi\Vert\omega)
  =(\alpha-1)\widetilde D_\alpha(\psi\Vert\omega)
  \label{eq:discussion-order-convexity}
\end{equation}
is convex on intervals above one where it is finite
\cite{Berta:2016vnw}. Related estimates establish monotonicity
of the SRD in $\alpha$ and transfer contractivity between
exponents, a useful step in proofs of data processing
\cite{Beigi:2013teh,Jencova:2016tqz}. 

Convexity under displacement
of a cut requires additional information about the physical
deformation. For the R\'enyi QNEC, that information can come from the
canonical translations of a half-sided modular inclusion.
The proof at integer orders $n\geq2$ represents these
translations as a positive contraction semigroup on Haagerup
$L^n(M)$. Modular covariance then supplies the identities needed
to establish convexity of the SRD along the semigroup
\cite{Kibe:2026wsg}. The order-one limit of our coherent-state
example recovers relative entropy, whose second cut derivative
equals $2\pi$ times the local coherent-state energy density.
This connects the calculation
to earlier results on relative entropy and its response to
null translations
\cite{Longo:2020CurrentEntropy,Ciolli:2020wave}. A broader
treatment of relative entropy for coherent excitations of
quasifree states on CCR algebras is given in
\cite{Bostelmann:2022coherent}.

Haagerup spaces and related weighted $L^p$ formulations also
support definitions of distinguishability and proofs of data
processing for general von Neumann algebras
\cite{Berta:2016vnw,Jencova:2016tqz,Jencova:2017txf}.
For $\alpha>1$ and finite SRD, equality in data processing
for a quantum channel characterizes reversibility for the
pair of states being compared \cite{Jencova:2016tqz}.

Quantum hypothesis testing gives another operational
interpretation of these quantities. Haagerup $L^p$ methods
enter the extension of discrimination bounds and Stein
asymptotics to general von Neumann algebras
\cite{Pautrat:2020wci}. The SRDs with $\alpha>1$ determine
the strong-converse exponent in discrimination of many
independent copies of two states. If the probability of falsely
accepting one state is required to decay at an exponential rate
exceeding the relative entropy, the probability of correctly
accepting it must also vanish. The strong-converse exponent
gives its optimal exponential decay rate. This characterization
was established for injective von Neumann algebras
\cite{Hiai:2021slp} and subsequently extended to general
von Neumann algebras \cite{Junge:2025kuz}.

Crossed products have a further role in quantum gravity.
In certain semiclassical and holographic constructions, a
Type~$\mathrm{II}_\infty$ crossed product describes a
gravitational observable algebra and admits an entropy
defined using its semifinite trace
\cite{Witten:2021unn,Chandrasekaran:2022eqq,Jensen:2023yxy}.
The explicit constructions in these notes help make this
algebraic structure concrete, but the distinction between
the two integration functionals remains essential:
a normalized Haagerup density satisfies
$\operatorname{tr}(h_\psi)=1$, while $\tau(h_\psi)=+\infty$.
It therefore does not itself define a normalized density
for the core trace. Connecting the Haagerup calculus to
gravitational entropy requires a physical identification
of the observable algebra and its states, together with
the assumptions on modular flow appropriate to the region
under consideration.

Beyond QFT and quantum information, noncommutative $L^p$
spaces are basic tools in quantum probability and
operator-algebraic analysis. They allow conditional
expectations, martingales, and ergodic averages to be studied
when the relevant state is not tracial \cite{Pisier:2003dsk}.
The reduction method of Haagerup, Junge, and Xu transfers
a broad class of problems from general noncommutative $L^p$
spaces to finite tracial algebras, with applications to
martingale and ergodic inequalities \cite{Haagerup:2008arm}.
These methods provide control of fluctuations and long-time
averages in noncommutative probability.

\begin{acknowledgments}
These notes developed out of discussions with Leandro Martinek, and a series of lectures delivered by T.K. to the Little Group at Instituto Balseiro. 
T.K thanks the students who attended the lectures, whose questions helped shape these notes into a better form.
The authors used LLMs as editorial aids for drafting and refining parts of the manuscript, and preparing the figure. We take full responsibility for its contents.
T.K. is supported by a Simons Foundation fellowship through the Targeted Grant to Instituto Balseiro. The work of P.R. has been supported by the Polish National Science Centre through Sonata grant (2022/47/D/ST2/02058). 
\end{acknowledgments}

\appendix

\section{Regular and untwisted realizations of the matrix core}
\label{app:matrix-charts}

We relate the regular crossed-product representation to the
matrix-field realization used in
Section~\ref{subsec:matrix-untwisted-core}.
Write $D=\e^{-K}>0$, with
$K=\operatorname{diag}(k_1,\ldots,k_n)$ and
$\sigma_r^D(a)=D^{ir}aD^{-ir}$; setting $D=\varphi$ recovers
the notation of the main text.

On $L^2(\mathbb R_s;\mathbb C^n)$, the regular generators are
\begin{equation}
  (\pi_{\mathrm{reg}}(a)\xi)(s)=D^{-is}aD^{is}\xi(s),
  \qquad
  (\lambda_{\mathrm{reg}}(r)\xi)(s)=\xi(s-r).
  \label{eq:app-matrix-regular-generators}
\end{equation}
With the Fourier convention
$(\mathcal F\xi)(t)=(2\pi)^{-1/2}
 \int_{\mathbb R}\e^{-its}\xi(s)\dd s$, these become
\begin{align}
  (\mathcal F\pi_{\mathrm{reg}}(a)\mathcal F^*
       \widehat\xi)_\alpha(t)
    &=\sum_\beta a_{\alpha\beta}
       \widehat\xi_\beta(t+k_\beta-k_\alpha),
  \label{eq:app-matrix-Fourier-pi}\\
  (\mathcal F\lambda_{\mathrm{reg}}(r)\mathcal F^*
       \widehat\xi)(t)
    &=\e^{-irt}\widehat\xi(t).
  \label{eq:app-matrix-Fourier-lambda}
\end{align}
The component translations are removed by the unitary
\begin{equation}
  (U\widehat\xi)_\alpha(t)=\widehat\xi_\alpha(t+k_\alpha).
  \label{eq:app-matrix-component-shift}
\end{equation}
Direct substitution gives
\begin{equation}
  U\mathcal F\pi_{\mathrm{reg}}(a)\mathcal F^*U^*=M_a,
  \qquad
  U\mathcal F\lambda_{\mathrm{reg}}(r)\mathcal F^*U^*
     =M_{\e^{-irt}D^{ir}},
  \label{eq:app-matrix-main-generators}
\end{equation}
where $M_A$ denotes multiplication by the matrix field $A(t)$.
These are the generators of
\eqref{eq:matrix-untwisted-pi}--\eqref{eq:matrix-untwisted-lambda}.

\bibliographystyle{JHEP}
\bibliography{refs}

\end{document}